\documentclass[aps,prx,twocolumn,superscriptaddress]{revtex4-2}
\usepackage[english]{babel}
\usepackage{graphicx}
\usepackage{dcolumn}
\usepackage{bm}
\usepackage{multirow}
\usepackage{amsmath,amssymb,amsfonts}
\usepackage{amsthm}
\usepackage{mathrsfs}
\usepackage[title]{appendix}
\usepackage{xcolor}
\usepackage{textcomp}
\usepackage{booktabs}
\usepackage{algorithm}
\usepackage{algorithmicx}
\usepackage{algpseudocode}
\usepackage{listings}
\usepackage{physics}

\newcommand{\be}{\begin{equation}}
\newcommand{\ee}{\end{equation}}

\newcommand{\ourtitle}{Realization of Two-dimensional Discrete Time Crystals\\ with Anisotropic Heisenberg Coupling}
\begin{document}
\title{\ourtitle}

\author{Eric D. Switzer}
\email{eric.switzer@nist.gov}
\affiliation{Donostia International Physics Center (DIPC), 20018 Donostia-San Sebastián, Euskadi, Spain}
\affiliation{Nanoscale Device Characterization Division, National Institute of Standards and Technology, Gaithersburg, Maryland 20899, USA}
\affiliation{Department of Physics, University of Central Florida, Orlando, Florida 32816, USA}

\author{Niall Robertson}
\affiliation{IBM Quantum, IBM Research Europe - Dublin, IBM Technology Campus, Dublin 15, Ireland}

\author{Nathan Keenan}
\affiliation{Department of Physics, Trinity College Dublin, Dublin 2, Ireland}
\affiliation{IBM Quantum, IBM Research Europe - Dublin, IBM Technology Campus, Dublin 15, Ireland}
\affiliation{Trinity Quantum Alliance, Unit 16, Trinity Technology and Enterprise Centre, Pearse Street, D02 YN67, Dublin 2, Ireland}

\author{Ángel Rodríguez}
\affiliation{Centro de Física de Materiales CFM/MPC (CSIC-UPV/EHU), 20018 Donostia-San Sebastián, Euskadi, Spain}

\author{Andrea D'Urbano}
\affiliation{IBM Quantum, IBM Research Europe - Dublin, IBM Technology Campus, Dublin 15, Ireland}

\author{Bibek Pokharel}\affiliation{IBM~Quantum,~IBM~T.J.~Watson~Research~Center,~Yorktown~Heights,~NY~10598,~USA}

\author{Talat S. Rahman}
\affiliation{Department of Physics, University of Central Florida, Orlando, Florida 32816, USA}
\affiliation{Donostia International Physics Center (DIPC), 20018 Donostia-San Sebastián, Euskadi, Spain}

\author{Oles Shtanko}
\email{oles.shtanko@ibm.com}
\affiliation{IBM Quantum, IBM Research -- Almaden, San Jose, California 35120, USA}

\author{Sergiy Zhuk}
\email{sergiy.zhuk@ie.ibm.com}
\affiliation{IBM Quantum, IBM Research Europe - Dublin, IBM Technology Campus, Dublin 15, Ireland}

\author{Nicolás Lorente}
\email{nicolas.lorente@ehu.eus}
\affiliation{Donostia International Physics Center (DIPC), 20018 Donostia-San Sebastián, Euskadi, Spain}
\affiliation{Centro de Física de Materiales CFM/MPC (CSIC-UPV/EHU), 20018 Donostia-San Sebastián, Euskadi, Spain}
\maketitle

\textbf{
A discrete time crystal (DTC) is the paradigmatic example of a phase of matter that occurs exclusively in systems out of equilibrium~\cite{khemani2016phase,else2016floquet,zaletel2023colloquium,Greilich_2024}. 
This phenomenon is characterized by the spontaneous symmetry breaking of discrete time-translation and provides a rich playground to study a fundamental question in statistical physics: what mechanism allows for \textit{driven} quantum systems to exhibit emergent behavior that deviates from their counterparts with time-independent evolution \cite{Polkovnikov_2011,Rudner_2013}? 
Unlike equilibrium phases, DTCs exhibit macroscopic manifestations of coherent quantum dynamics, challenging the conventional narrative that thermodynamic behavior universally erases quantum signatures~\cite{Zurek_2003}. 
However, due to the difficulty of simulating these systems with either classical or quantum computers, previous studies have been limited to a set of models with Ising-like couplings - and mostly only in one dimension~\cite{ippoliti_many-body_2021,mi2022time,frey2022realization} - thus precluding our understanding of the existence (or not) of DTCs in models with interactions that closely align with what occurs in nature~\cite{Kimura_2007,Breunig_2013,Toskovic_2016,Faure_2018}. 
In this work, by combining the latest generation of IBM quantum processors with state-of-the-art tensor network methods, we are able to demonstrate the existence of a DTC in a two-dimensional system governed by anisotropic Heisenberg interactions. 
Our comprehensive analysis reveals a rich phase diagram encompassing spin-glass, ergodic, and time-crystalline phases, highlighting the tunability of these phases through multiple control parameters. 
Crucially, our results emphasize the interplay of initialization, interaction anisotropy, and driving protocols in stabilizing the DTC phase. 
By extending the study of Floquet matter beyond simplified models, we lay the groundwork for exploring how driven systems bridge the gap between quantum coherence and emergent non-equilibrium thermodynamics~\cite{Davoudi_2024}.
}

The mechanisms by which the laws of thermodynamics arise from the underlying unitary evolution of quantum mechanics is a topic of fundamental interest in the physical sciences. 
For example, it is now well understood that certain out-of-equilibrium quantum systems can indeed thermalize and hence act as their own heat bath such that local observables converge to thermal values described by a Gibbs ensemble \cite{srednicki1994chaos, rigol2008thermalization} or a generalized Gibbs ensemble \cite{rigol2007relaxation, vidmar2016generalized}. 
These out-of-equilibrium phenomena are typically addressed by considering toy-models in the context of a ‘quantum quench’, whereby non-trivial dynamics are induced by a single rapid change of the Hamiltonian of the model. 
This immediately raises the question of what might be different for the thermodynamic behavior of quantum systems outside of the quench paradigm such as \textit{driven} quantum systems in which energy is constantly exchanged with the environment -  a situation which may more closely reflect naturally occurring thermodynamic processes. 
A striking example of the fundamentally different nature of quantum systems with and without a drive is provided by the emergence of discrete time crystals (DTC) \cite{khemani2016phase,else2016floquet,von2016absolute,zaletel2023colloquium}. 

Discrete time crystals emerge in certain periodically driven many-body-localized (MBL) spin systems \cite{ponte2015many-body,abanin2019many-body}, where periodic driving combines with stable long-range ordering.
When the drive (nearly) flips every spin, the ordering imposes a structure on the single-period unitary evolution operator, resulting in pairs of eigenstates separated by an exact $\pi$-phase rotation that remains stable against small perturbations \cite{khemani2016phase,else2016floquet,von2016absolute}.
Consequently, local observables exhibit oscillations with a period that is a multiple of the driving frequency, persisting indefinitely in perfectly isolated systems.
This subharmonic response represents a spontaneous breaking of discrete time-translation symmetry, analogous to the breaking of continuous spatial symmetry in conventional solid-state crystals.

\begin{figure*}[t!]
    \centering
\includegraphics[width=\linewidth]{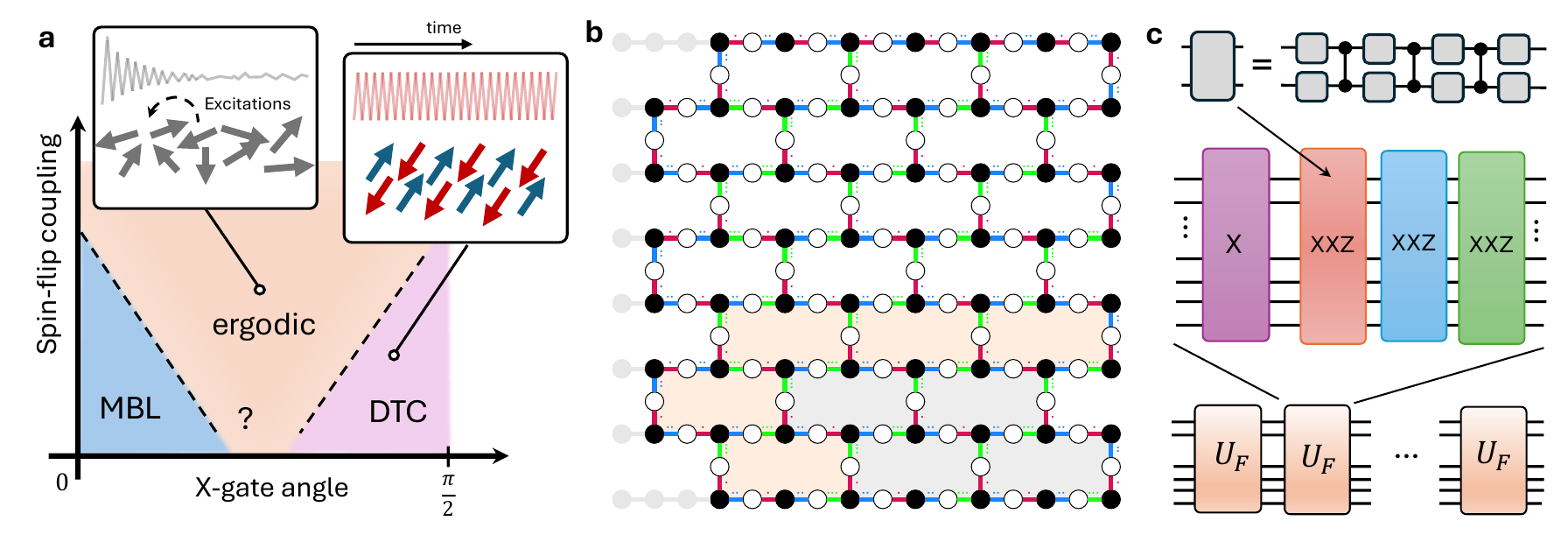}
    \caption{\textbf{Two-dimensional transitions driven by spin-flip coupling.}  \textbf{a}, Discrete time crystals (DTC) are stabilized through a long-range order induced by disordered Ising couplings. Introducing XY spin-flip couplings acts as an additional mechanism for state-dependent thermalization. However, the stability of time-crystal ordering in two dimensions in the presence of such coupling, along with the precise location of phase boundaries, remains an open question. \textbf{b}, Circuit layout of the experimental 144-qubit system implemented on the \texttt{ibm\_fez} device. Circles represent qubits, while lines connecting them denote two-qubit gates. Qubit colors indicate their initial states, with the example shown corresponding to a N\'eel state. Colors and dot patterns on connections specify the sequence and order of two-qubit gates, as detailed in the rightmost panel. \textbf{c}, Each two-qubit gate is transpiled into single-qubit gates and three controlled-Z operations, which are native to the \texttt{ibm\_fez} device. Two-qubit gates are arranged in three layers to cover entire lattice workspace forming a Floquet cycle that repeats $t$ times.  
}  \label{fig:main_scheme}
\end{figure*}

In this work, we explore a new setup for observing discrete time crystals. 
Previously, most studies focused on one-dimensional Ising-type couplings between spins due to their lower classical simulation complexity and straightforward implementation in experimental setups, which often rely on linear arrays \cite{choi2017observation,zhang2017observation,pal_temporal_2018,rovny_observation_2018,rovny_31mathrmp_2018,smits_observation_2018,autti_observation_2018,randall2021many,kongkhambut_observation_2022,mi2022time,frey2022realization,liu_higher-order_2024}.
Building on prior work, we extend the experimental investigation to a two-dimensional system on a ``decorated'' (heavy) hexagonal lattice \cite{shinjo2024unveiling,fernandes2024bipartite} and introduce spin-flip coupling. 
Exploring such Heisenberg-type models is motivated by their applicability to several physical systems. 
Heisenberg-type interactions naturally arise in a wide range of physical systems, from single-molecule magnets and metallic chains to quantum dot-based architectures~\cite{li_discrete_2020,throckmorton_effects_2022,sarkar_time_2024,Greilich_2024,shukla_prethermal_2024}. 
The ability to realize such models in different experimental settings underscores the broader relevance of our findings beyond the specific implementation on superconducting qubits.

Our quantum simulations demonstrate that, despite weaker localization and theoretical arguments challenging the existence of MBL in two dimensions \cite{deroeck2017stability}, disordered systems exhibit resilience to weak spin-flip coupling across various initial states, at least for short times.
At the same time, such strong coupling drives a transition into an intermediate ergodic phase similar to that in one-dimensional systems \cite{ippoliti_many-body_2021,sahay2021emergent}, see Fig.~\ref{fig:main_scheme}a. 
Moreover, unlike the Ising model, we observe a striking contrast between initial states on which spin-flip interactions act differently. 
As an example, we consider the N\'eel state, which maximizes spin-flip uncertainty, and a polarized state, an eigenstate of the spin-flip interaction.
Oscillations originating from the N\'eel state decay rapidly, while the fully polarized state demonstrates a stable subharmonic response that resembles the behavior of quantum many-body scars \cite{turner2018weak,maskara_discrete_2021} including cat-scar DTCs \cite{huang_analytical_2023, bao_creating_2024}.

\textbf{Model and its implementation}. Our model, illustrated in Fig.~\ref{fig:main_scheme}c, describes a discrete-time evolution. 
The state of the system is expressed as $|\psi_t\rangle = U_{F}^t|\psi_0\rangle$, where $t$ represents integer Floquet cycles.  
Here, $|\psi_0\rangle$ denotes a product state of spins, and $U_{F}$ is the Floquet unitary operator given by 
    \be\label{eq:UF_def}
    U_{F} = U_{\rm XXZ}^{(3)}U_{\rm XXZ}^{(2)}U_{\rm XXZ}^{(1)}U_{\rm X},
    \ee
expressed in terms of three sub-cycle layers $U_{\rm XXZ}^{(k)}$ and an X-gate rotation $U_{\rm X}$ as described below (see Fig.~\ref{fig:main_scheme}b for schematics).

The X-gate rotation corresponds to a periodic transverse kick pulse, expressed as  
    \be\label{eq:U_x_def}
    U_{\rm X} = \prod_{i=1}^N \exp(-i\phi X_{i}),
    \ee  
parameterized by the X-gate 
angle $\phi$. 
The cases $\phi = 0$ and $\phi = \pi/2$ correspond to specific cases of no kick and a perfect $X$-flip of all qubits, respectively.  

\begin{figure*}[t!]
    \centering
\includegraphics[width=\textwidth]{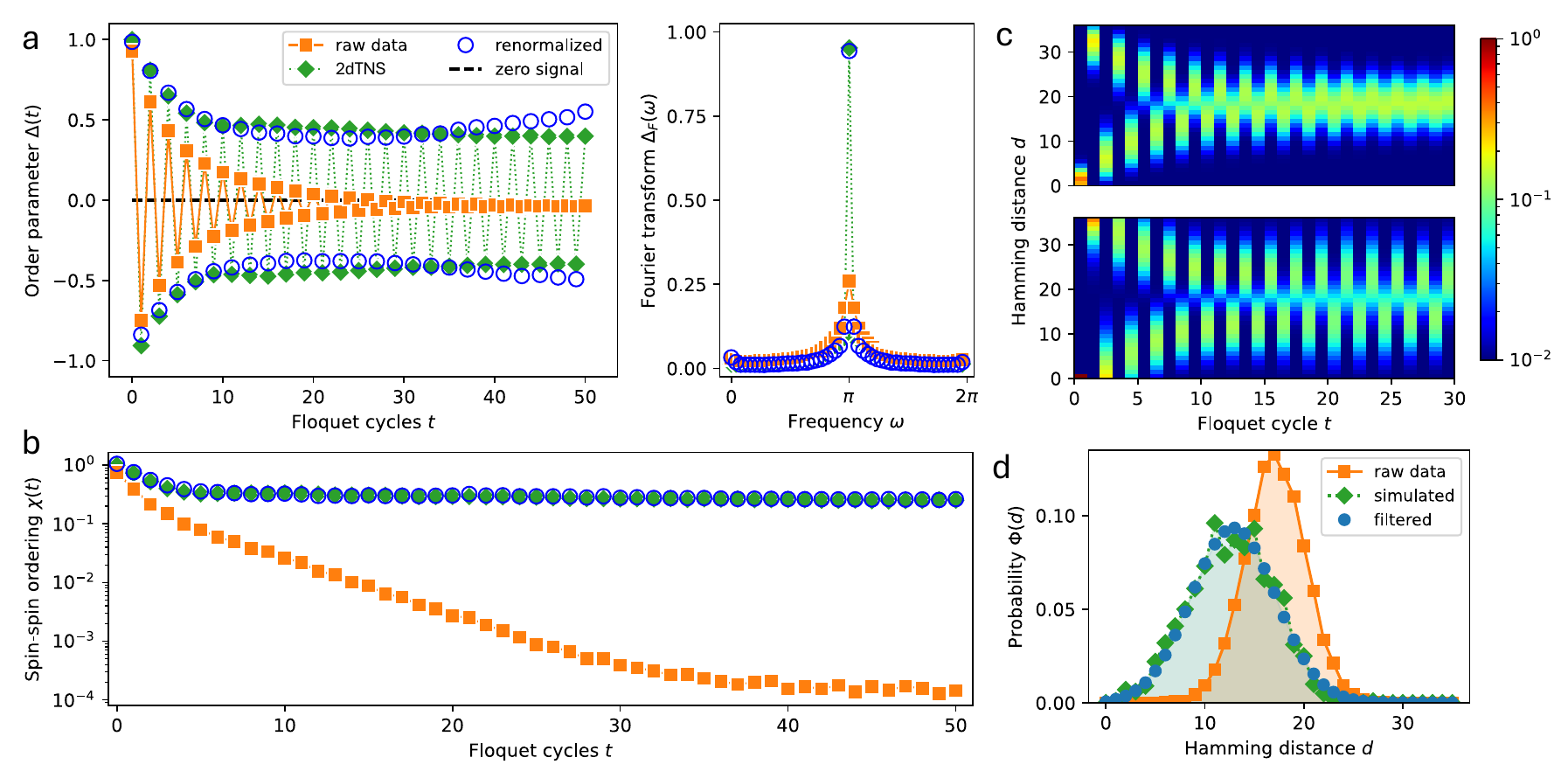}
    \caption{\textbf{Subharmonic behavior.} Performance of the device in the discrete time-crystalline regime. The results are compared to numerical simulations, where we use a two-dimensional tensor network state (2dTNS), see Section~\ref{sec:supp_mat_2dtns} of the Supplementary Information. Noise recovery is performed using the renormalization methods described in Sections~\ref{sec:supp_methods_signalrecov} and \ref{sec:supp_signalrecov} of the Supplementary Information. \textbf{a}, The order parameter $\Delta(t)$, defined in Eq.~\eqref{delta_parameter}, is plotted as a function of the number of Floquet cycles, illustrating the subharmonic response characteristic of the discrete time-crystalline phase. The data shown include raw device measurements (orange squares), classical simulations (green diamonds), and noise-recovered data (empty circles). The right panel presents the Fourier transform of the order parameter. \textbf{b}, The spin-spin correlation parameter $\chi(t)$, defined in Eq.~\eqref{eq:spin_glass_order}, is shown. The notations are consistent with panel \textbf{a}. \textbf{c}, The evolution of Hamming distance distributions, defined in Eq.~\eqref{eq:distr_ham_dist}, is presented on a logarithmic scale for raw data (top) and mitigated data (bottom) in a $2 \times 2$ system. The mitigated distribution converges to a steady-state value. \textbf{d}, An illustration of noise filtering through a comparison of Hamming distance distributions for raw device data (orange squares), 2dTNS classical simulations (green diamonds), and noise-filtered results (blue circles), exemplified at Floquet cycle 25, revealing small skewness and sizable broadening.
}
    \label{fig:fig2}
\end{figure*}

The sub-cycle coupling layers are constructed as a composition of non-overlapping two-qubit gates, expressed as 
    \be\label{eq:xxz_layers}
    U_{\rm XXZ}^{(k)} = \prod_{(i,j)\in G_{k}} U_{ij},
    \ee  
where $G_k$ represents a list of qubit pairs subjected to gates at sub-layer $k$, as illustrated in Fig.~\ref{fig:main_scheme}c. 
The coupling gate is an implementation of a two-qubit disordered XXZ model in the form
    \be\label{eq:U_ij_def}
    U_{ij} = \exp\Bigl[-iJ_{ij}(\epsilon X_{i}X_{j}+\epsilon Y_{i}Y_{j}+Z_{i}Z_{j})\Bigl],
    \ee
where $X_{i}$, $Y_{i}$, and $Z_{i}$ are Pauli operators for qubit $i$, $J_{ij} = 1+\delta_{ij}$ denotes the coupling strength, $\delta_{ij} \in[-0.5,0.5]$ represents uniformly sampled disorder, and $\epsilon$ is the spin-flip coupling strength. 
The disorder in the coupling induces a many-body localization regime, which prevents the system from absorbing energy from the periodic drive. 
As a result, the system avoids reaching an ``infinite-temperature'' state quickly, characterized by trivial values for all local observables. 
Notably, we found that using a single disorder realization is sufficient for describing sums of single-qubit observables, given the large number of qubits employed.
In particular, we use $N=144$ qubits in two initial states: a N\'eel state, illustrated in Fig.~\ref{fig:main_scheme}b, and a fully polarized state (results shown only in Fig.~\ref{fig:polarized_state}).

To implement this model, we employ the 156-qubit Heron r2 quantum processor, \texttt{ibm\_fez}, which uses fixed-frequency transmon qubits. 
From this device, we select a 144-qubit subset that includes a decorated lattice of $3\times 7$ heavy hexagons, along with smaller configurations of $2\times 2$ and $3\times 3$ containing 35 and 68 qubits, respectively, as illustrated in Fig.~\ref{fig:main_scheme}b. These smaller subsystems are used to examine the scaling of order parameters with system size and to mitigate the effects of noise.

The quantum dynamics is implemented through two sets of runs: one consisting of 50 Floquet cycles with 20,000 shots per circuit, and another comprising 30 Floquet cycles with 5,000 shots per circuit. 
For all experiments, the two-qubit (2Q) gate depth per cycle is 9, which is required to realize overlapping unitary operators in Eq.~$\eqref{eq:U_ij_def}$ sequentially. 
Consequently, the maximum circuit depth is 450 for 50-cycle experiments and 270 for 30-cycle experiments. 
Further details regarding the device architecture and implementation can be found in Supplementary Information, Sections~\ref{sec:supp_methods_implementation} and \ref{sec:supp_device}. 
The signal was processed using a renormalization technique that accounts for the effects of depolarization and amplitude damping. 
For further details, refer to Supplementary Information, Sections~\ref{sec:supp_methods} and \ref{sec:supp_signalrecov}.

To verify the results of the quantum hardware, we use classical simulations based on tensor networks \cite{paeckel2019time}. 
We consider two families of methods: Matrix Product States (MPS) and two-dimensional tensor network states. 
MPS-based methods are particularly well-suited to the study of one-dimensional models with local interactions. 
These methods also benefit from efficient contraction algorithms to calculate expectation values as well as schemes for
optimal truncation of the bond dimension. 
However, to apply MPS-based methods to models in two dimensions, one must ``unroll'' the two-dimensional model into its one-dimensional representation. 
To go beyond these limitations, we use two-dimensional tensor network states (2dTNS) \cite{verstraete2004renormalization} with a topology that matches our lattice, thus maintaining the locality of the model's interactions and keeping the required bond dimension low. 
However, unlike MPS-based methods, exact contraction of a two-dimensional network is not efficient \cite{schuch2007computational} and truncations cannot be performed optimally. 
Instead, we use approximate methods, such as belief propagation \cite{tindall2023gauging}. 
To control the error of such approximations, we compare their results with MPS, when possible. 

\begin{figure*}[t!]
    \centering
\includegraphics[width=1.0\textwidth]{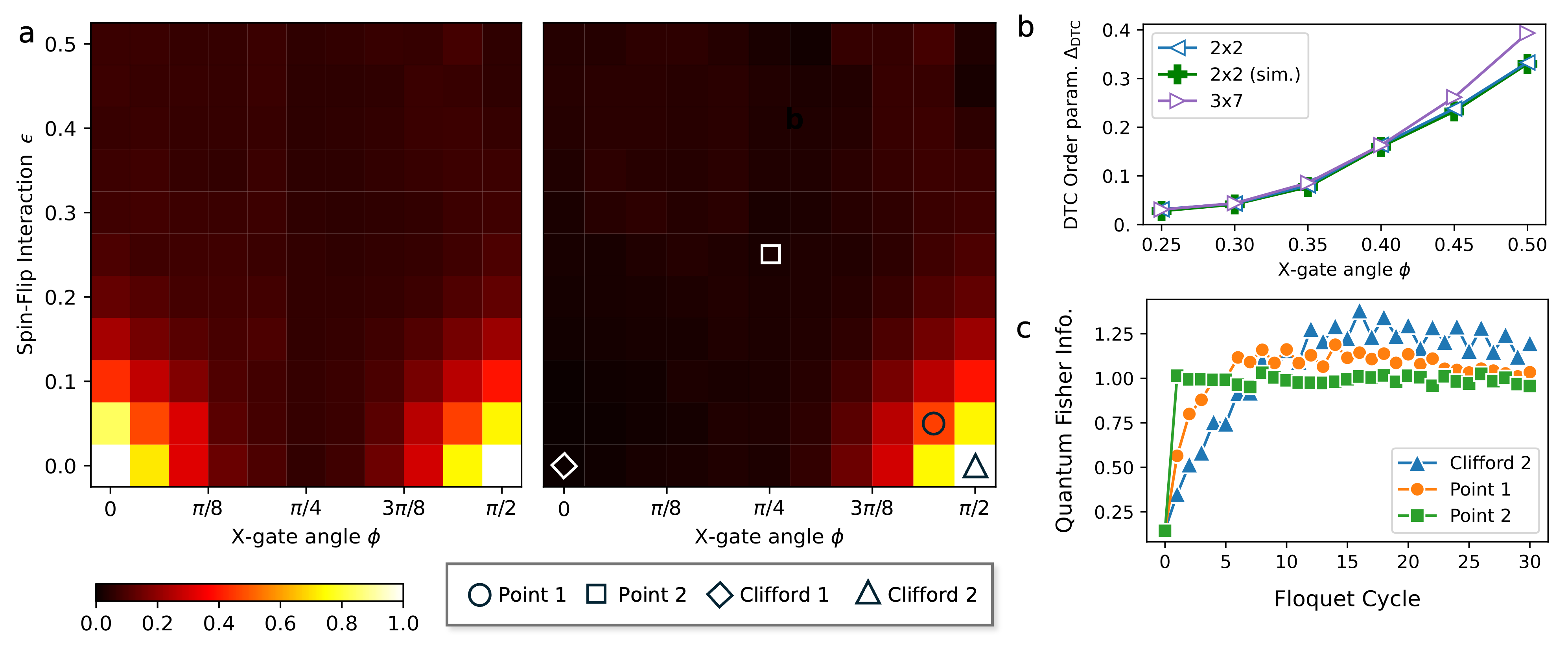}
    \caption{\textbf{Exploring the parameter space.}  \textbf{a}, Dependence of the MBL order parameter [Eq.~\eqref{eq:MBL}] (left) and the DTC order parameter [Eq.~\eqref{eq:DTC}] (right) on the spin-flip interaction strength $\epsilon$ and X-gate rotation angle $\phi$ for $T=30$ Floquet cycles. Regions where the order parameters approach unity correspond to crossovers into the glassy localized regime (left) and the discrete time crystal (DTC) regime (right). Key reference points include: the integrable MBL Clifford point (Clifford 1), the integrable DTC Clifford point (Clifford 2), the non-integrable DTC point (Point 1), and the ergodic point (Point 2).  \textbf{b}, Scaling behavior of the DTC order parameter with increasing system size at $\epsilon = 0.1$, suggesting a potential crossover to the DTC phase. White dots represent normalized hardware results for $2\times 2$ and $3\times 7$ systems, while filled points correspond to classical simulations for the $2\times 2$ system.  \textbf{c}, Quantum Fisher information for different reference points identified in panel a, showcasing the starkly different time dependence of entanglement growth between the localized DTC phase (Clifford 2, blue triangles, and Point 1, orange circles) and the ergodic phase (Point 2, green squares).
}
    \label{fig:phase_diagr}
\end{figure*}

\textbf{Results.} Similar to one-dimensional systems \cite{ippoliti_many-body_2021}, we anticipate probing transitions between three distinct phases: an ordered localized spin-glass phase, a discrete time crystal phase, and a featureless ergodic phase. 
To differentiate these phases, we define specific observables. 
A defining characteristic of the discrete time crystal phase is the appearance of persistent, period-doubled oscillations in spin dynamics, accompanied by long-range spatial order. 
This behavior is quantified through the evolution of the time correlation function of the same qubit. 

Consider an initial product state in the computational basis, i.e., 
    $
    |\psi_0\rangle = |s_1\rangle \otimes \dots \otimes |s_N\rangle,
    $ 
where $s_i \in \{-1,1\}$ specifies the spin projection along the $z$-axis. Then the time correlation function takes the form  
    \be\label{delta_parameter}
        \Delta(t) = \frac 1N\sum_{i=1}^N\langle\psi_0| Z_i(0)Z_i(t)|\psi_0\rangle= \frac 1N\sum_{i=1}^N s_i \langle \psi_t|Z_i|\psi_t\rangle,
    \ee
where $N$ is the total number of qubits, the integer $t$ denotes the number of Floquet cycles and the operator $Z_i(t) = U_F^{t\dag} Z_i U_F^t$ represents the Pauli-$Z$ operator in the Heisenberg picture. 
This measure effectively quantifies the individual spin memory of the initial state that persists over time.  

The results of the time-correlation measurements in the time-crystalline regime (specifically at the point $\phi = 0.45\pi$, $\epsilon = 0.05$) are presented in Fig.~\ref{fig:fig2}a. 
The measured data exhibit distinct double-periodic oscillations, which gradually decay over time due to hardware noise, in contrast to the idealized classical simulations. 
The presence of double-periodic oscillations is further corroborated by the Fourier transform plot shown in the right panel of Fig.~\ref{fig:fig2}a, which reveals a pronounced peak at the frequency $\omega = \pi$. 
To mitigate the effects of noise, we employ the renormalization technique described in Eq.~\eqref{eq:renorm} to recover the noiseless value of $\Delta(t)$. 
Specifically, parameters $c(t)$ and $c'(t)$ in Eq.~\eqref{eq:renorm} were extracted from data obtained on a $3 \times 3$ system and subsequently applied to the full $3 \times 7$ system. 
The recovery of the original peak is also apparent in the frequency components.  
 
In addition to examining time-correlation functions, we also analyze spatial correlations. We define a same-time spin correlation order parameter
    \be\label{eq:spin_glass_order}
        \chi(t) = \frac{1}{M} \sum_{\langle i,j \rangle} \langle \psi_t| Z_i Z_j|\psi_t \rangle^2,
    \ee  
and the sum is taken over all $M$ nearest neighbors.
The results are shown in Fig.~\ref{fig:fig2}b. 
Similar to the time correlations, the nearest-neighbor correlations decay over time due to hardware noise. This behavior is demonstrated by a comparison with the noiseless result obtained from classical simulations. 
To recover the noiseless values for the $3 \times 7$ system, we apply a renormalization technique analogous to that described in Eq.~\eqref{eq:renorm}. This approach uses noise parameters extracted from a smaller $3 \times 3$ system (see Supplementary Information, Section~\ref{sec:supp_signalrecov_correlations} for details). 
The results demonstrate that the system retains spin-spin ordering over time, which is a key feature for the emergence of the ordered phase.  
We also find that by expanding the set $M$ to all site pairs (which is equivalent to determining the Edwards-Anderson spin glass order parameter \cite{mi2022time}), the noisy hardware results are similar in behavior to the results of Fig.~\ref{fig:fig2}b, see Supplementary Information, Section~\ref{sec:supp_spinglass} for more details.

In addition to low-order correlations, it is instructive to examine a more complex measure that characterizes the circuit output. 
One such metric is the Hamming distance, which quantifies the minimum number of bit flips required to transform the measured bit string $\{s'_i(t) = \langle \psi_t|Z_i|\psi_t\rangle\}$ into the initial configuration $\{s_i\}$ 
The distribution of Hamming distances $\Phi(d,t)$ can be employed to characterize various correlations, and the variance of the distribution corresponds to the quantum Fisher information (QFI). 
The full distribution $\Phi(d)$ is obtained by sampling the bit strings $\{s'_i\}$ with appropriate probabilities, a process that is carried out using quantum hardware, see Supplementary Information, Section~\ref{sec:supp_signalrecov_hamming} for more details. 

The results for the Hamming distance distribution for a $2 \times 2$ system are presented in Fig.~\ref{fig:fig2}c-d. 
The measured distribution is visibly influenced by hardware noise. 
We observe that this effect can be effectively modeled as a process in which the bit strings undergo random and independent flips with a probability $p(t)$ that depends on the circuit cycle depth $t$. 
Using this approximation, we recover the original distributions conditioning the recovery on their mean and variance, which correspond to the classically simulated values of the time correlation and QFI (see Supplementary Information, Section~~\ref{sec:supp_signalrecov_hamming} for more details). 
The experiment demonstrates that the Hamming distance rapidly converges to two distinct, non-thermal distributions at odd and even periods. 
The shapes of these distributions are determined by the localization properties of the system. 
Unlike in the one-dimensional Ising model \cite{ippoliti_many-body_2021}, the distributions are nearly Gaussian, reflecting a higher degree of thermalization and exhibiting only a small degree of skewness.

\begin{figure}[t!]
    \centering
\includegraphics[width=0.9\linewidth]{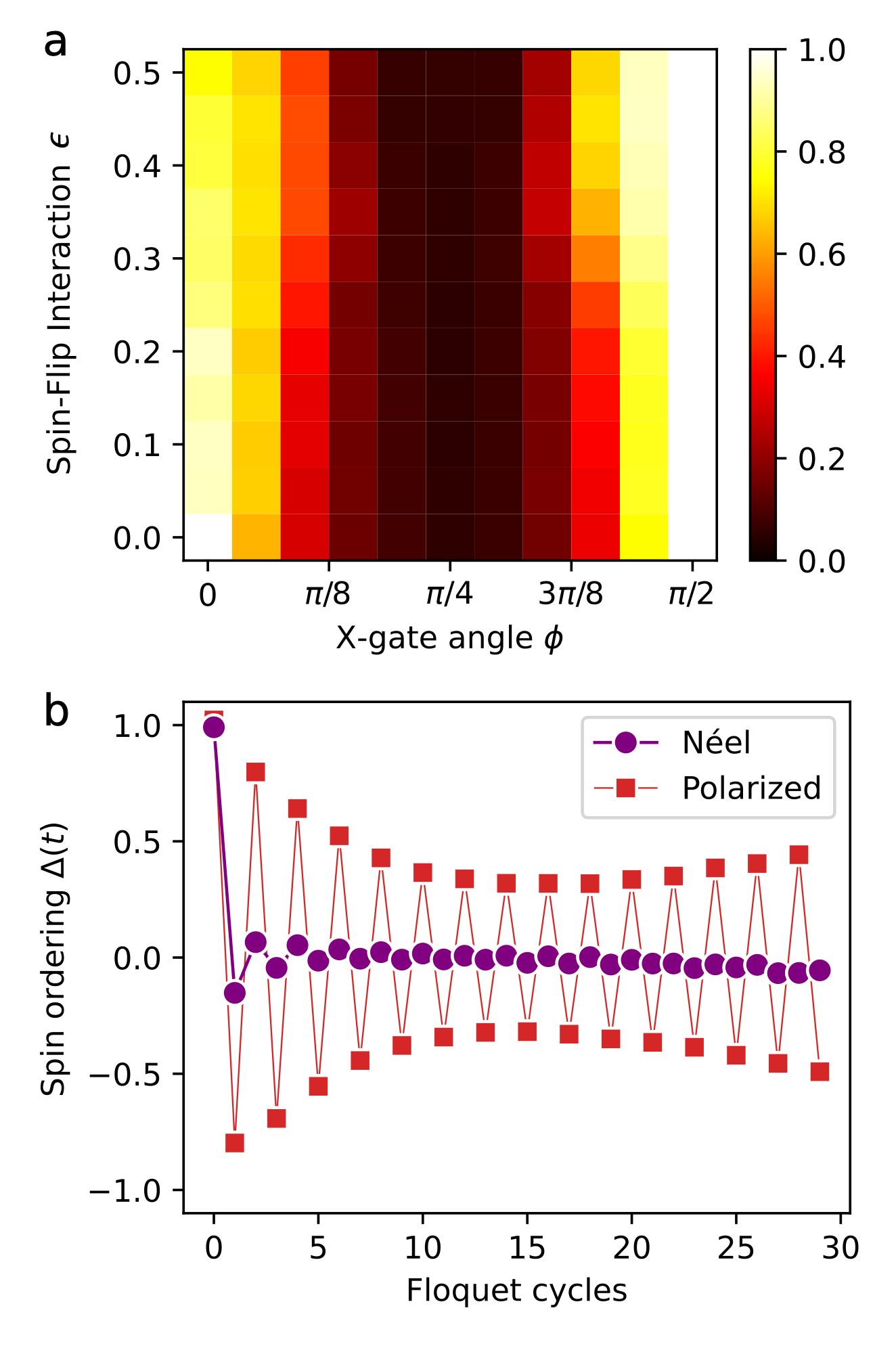}
    \caption{\textbf{Prethermal scar-like behavior.} Data corresponding to the polarized initial state $|\psi\rangle = \ket{\uparrow}^{\otimes N}$. \textbf{a}, The MBL order parameter as a function of spin-flip interaction strength $\epsilon$ and X-gate angle $\phi$. In contrast to Fig.~\ref{fig:phase_diagr}a, the order parameter remains substantially large across a wide range of $\epsilon$, indicating enhanced stability for the polarized state. \textbf{b}, A comparison of the spin ordering $\Delta(t)$ as a function of Floquet cycles $t$ for the polarized and N\'eel initial states at $\epsilon = 0.25$ and $\phi = 0.4\pi$. The polarized state does not exhibit decay and shows robust oscillations, highlighting its prethermal behavior.}
  \label{fig:polarized_state}
\end{figure}

The quantity defined in Eq.~\eqref{delta_parameter} enables us to construct order parameters that distinguish existing regimes of the system as we vary the parameters $\epsilon$ and $\phi$ in the model specified by Eqs.~\eqref{eq:U_x_def} and \eqref{eq:U_ij_def}. 
Recent work \cite{fernandes2024bipartite} has explored classical simulations of this problem for the special case $\epsilon = 0$. 
For nonzero values of $\epsilon$, quantum correlations significantly increase the complexity of the problem (see Supplementary Information, Section~\ref{sec:supp_signalrecov_correlations} for further details). 
To characterize the phases, we define two order parameters:  
    \begin{subequations}
        \begin{align}
        \Delta_{\rm MBL} &= \frac 1{NT}\sum_{t=0}^{T} \Biggl|\sum_{i=1}^N s_i \langle Z_i(t)\rangle\Biggl|, \label{eq:MBL} \\
        \Delta_{\rm DTC} &= \frac 1{NT}\sum_{t=0}^T\sum_{i=1}^N (-1)^t s_i \langle Z_i(t)\rangle, \label{eq:DTC}
        \end{align}
    \end{subequations}
where $N$ is the total number of qubits and $T \gg 1$ represents the maximum cycle depth. 
The parameter $\Delta_{\rm MBL}$ reflects the stability of many-body localization. 
It vanishes in the ergodic phase, whereas both the trivial spin-glass and discrete-time crystal phases exhibit nonzero values for any $T$. 
In contrast, the parameter $\Delta_{\rm DTC}$ is expected to be nonzero only in the discrete-time crystalline phase. 
The results are presented in Fig.~\ref{fig:phase_diagr}a for $T=30$. 
The color plots reveal the presence of a glassy regime for small values of $\phi$ and a discrete-time crystalline regime for $\phi$ near $\pi/2$, both occurring at nonzero values of $\epsilon$. 
Notably, a comparison between the $2 \times 2$ and $3 \times 7$ systems demonstrates improved stability of the order parameters. This supports the existence of a transition between localized and ergodic phases, as shown in Fig.~\ref{fig:phase_diagr}b.  

Another key parameter for characterizing the phase transition is the quantum Fisher information (QFI). 
The logarithmic growth of QFI has previously been used as evidence for the existence of a many-body localized phase~\cite{Smith_2016,li_discrete_2020}.
The QFI values shown in Fig.\ref{fig:phase_diagr}c provide evidence of very slow growth of entanglement within the glassy and discrete-time crystal regimes. This behavior is depicted in Fig.~\ref{fig:phase_diagr}a.
In contrast, for points located in the center of the parameter space, we observe a faster rate of QFI growth, consistent with the behavior expected in an ergodic regime.  

All the results discussed above are based on the N\'eel initial state, as illustrated in Fig.~\ref{fig:main_scheme}a. 
We expect that most input states will exhibit behavior similar to that of the N\'eel state, with the notable exception of fully polarized states or states close to them.  
Polarized states are distinct because they are eigenstates of the spin-flip component of the Hamiltonian in Eq.~\eqref{eq:xxz_layers}.
Consequently, these states exhibit long-time prethermal behavior, even for $\epsilon \sim 1$.  
Fig.~\ref{fig:polarized_state} shows the renormalized hardware data for the polarized initial state, emphasizing the pronounced difference in the amplitude of discrete time crystal oscillations between polarized and N\'eel states. 
The observed stability of these oscillations closely resembles the behavior of quantum many-body scars \cite{turner2018weak,maskara_discrete_2021} and cat-scar DTCs \cite{huang_analytical_2023, bao_creating_2024}.

\textbf{Discussion}. The stability of the fully polarized state in the XXZ Hamiltonian under periodic transverse kicks raises questions about weak ergodicity breaking. 
The introduction of periodic kick pulses with generic phases (i.e. excluding $n\pi/2$ for integer $n$) breaks the U(1) symmetry of the XXZ Hamiltonian, potentially leading to thermalization through eigenstate mixing during Floquet evolution. 
Remarkably, the fully polarized state shows robustness near $\phi = 0$ and $\pi/2$, resulting in a phase diagram distinct from earlier findings \cite{ippoliti_many-body_2021}, and this stability may stem from stable state evolution trajectories that prevent eigenstate mixing \cite{maskara_discrete_2021}. 
Such trajectories, resembling those in cat scar DTCs, suggest a novel mechanism preserving magnetic order even in the presence of perturbations \cite{huang_analytical_2023,bao_creating_2024}.

So far we have considered the emergent thermodynamic behavior of a two-dimensional driven quantum system with anisotropic Heisenberg coupling.
We note that there are two sources of the drive in our model: the X-gate pulse $U_X$ in Eq.~\eqref{eq:U_x_def}, and the discrete time evolution of the Heisenberg model, i.e., $U_F$ in Eq.~\eqref{eq:UF_def}. 
An interesting question therefore naturally emerges: how might the phase diagram of the kicked-Heisenberg model change if one of these drives is turned off, i.e., the Heisenberg part of the evolution becomes a continuous time evolution operator?
To probe this, one can formally define a family of models characterized by a parameter $k$ which determines the number of discrete time steps used to implement one Floquet cycle of the Heisenberg evolution.
In this work, we have considered $k=1$, while the continuous time analog would correspond to $k=\infty$.
It remains an open question on how the results presented here could vary as a function of $k$.
With the quantum hardware challenges of probing larger $k$, one could instead turn to Multiproduct Formula-based methods \cite{robertson2024tensor, zhuk2023trotter, vazquez2023well} to study linear combinations of the results for small values of $k$, in order to gain insight to our model in Eq.~\eqref{eq:UF_def} with continuous time evolution.

\textbf{Conclusion}. This work presents the first experimental investigation of a two-dimensional discrete time crystal with anisotropic Heisenberg coupling. 
By developing scalable noise renormalization methods, we calculated the time evolution of a particular family of observables near the ergodic phase transition — a regime generally regarded as challenging for classical computation. 
This allowed us to construct a detailed phase diagram of the system and to detect the boundaries between three regimes: one ergodic and two localized. 
Our findings demonstrate that the interplay between disorder and periodic driving can sustain stable and robust quantum dynamics in two dimensions. 
Crucially, our findings establish that many-body localization mechanisms, despite their theoretical fragility in two dimensions, can provide sufficient protection against thermalization, allowing for stable subharmonic response in a driven quantum system.
Furthermore, we observed surprising scar-like behavior in the discrete time crystals, which, interestingly, is a phenomenon that appears to arise as a result of the different symmetry structure of the Heisenberg model as compared to the Ising model. 
This discovery provides a new perspective on weak ergodicity breaking in driven many-body systems, potentially connecting time crystals to quantum many-body scars and prethermal phases.
The methods proposed in this work open new avenues for exploring large quantum systems and the emergence of exotic phases of matter on current quantum hardware even in the presence of noise.

\textbf{Author Contributions.}
O.S. developed the theoretical framework for analyzing the quantum dynamics presented in this work. 
E.S. and O.S. designed the hardware implementation, and E.S. performed the hardware implementation. 
N.R. and N.K. performed classical simulations, O.S. and E.S. performed data postprocessing, O.S. and S.Z. designed and implemented signal recovery from noisy data, A.R. and B.P. contributed to performing the experiment, and A.D. contributed to classical simulations and processing the data. 
N.L., O.S., S.Z., E.S., and T.R. helped guide the experiment and interpret the data. 
The manuscript was written by O.S., E.S., N.R., S.Z, B.P., and N.L. 
All authors provided suggestions for the experiment, discussed the results, and contributed to the manuscript.

\textbf{Acknowledgments.} We thank Martin Mevissen, Sergey Bravyi, Antonio Mezzacapo, and Garnett W. Bryant for helpful discussions. 
E.S., A.R., and N.L. thank the Basque Government BasQ initiative for making this project possible. 
E.S. and T.R. acknowledge support in part from the Department of Energy, grant number DE-FG02-07ER46354. 
Any mention of equipment, instruments, software, or materials does not imply recommendation or endorsement by the National Institute of Standards and Technology.

\let\oldaddcontentsline\addcontentsline
\renewcommand{\addcontentsline}[3]{}
\bibliography{references}
\let\addcontentsline\oldaddcontentsline

\clearpage
\pagebreak
\setcounter{page}{1}
\setcounter{equation}{0}
\setcounter{figure}{0}
\renewcommand{\theequation}{S.\arabic{equation}}
\renewcommand{\thefigure}{S\arabic{figure}}
\renewcommand*{\thepage}{S\arabic{page}}
\onecolumngrid
\begin{center}
{\large \textbf{Supplementary Information for \\ ``\ourtitle"}}\\
\end{center}
\tableofcontents
\section{Methods}
\label{sec:supp_methods}
\subsection{Signal recovery from noisy data}
\label{sec:supp_methods_signalrecov}
The presence of hardware noise requires robust techniques for extracting signals, such as the expectation values of Pauli operators, from noisy measurements. 
While conventional error mitigation methods \cite{Wallman2017Twirling,Temme2017PECandZNE,Li2017ZNE,van2023probabilistic} offer reasonable improvements for short-time dynamics, they fail to recover signals in circuits deep enough to explore time-crystalline behavior within our framework. 
To address this limitation, we introduce a physics-inspired recovery procedure that leverages specific characteristics of the noisy dynamics. 
In particular, we show that the influence of noise on a collective observable $O$ can be effectively modeled as  
    \be
    \langle O\rangle_{\rm noisy} = f(t) \langle O\rangle + c(t),
    \ee
where $\langle \cdot \rangle_{\rm noisy}$ and $\langle \cdot \rangle$ denote noisy and noiseless expectation values, respectively. 
Here, $f(t)$ represents an amplitude attenuation factor that decays asymptotically, i.e., $f(t)\to 0$ as $t \to \infty$, while $c(t)$ denotes an offset capturing the observable's long-time behavior. 
We further assume that the offset exhibits a double-periodic structure, such that $c(t) = c(t+2)$.  

Assuming that $f(t)$ exhibits a weak dependence on the model parameters, the observable $\langle O\rangle$ can be recovered using the following expression:  
    \be\label{eq:renorm}
    \langle O\rangle = \langle O\rangle'\frac{\langle O\rangle_{\rm noisy}-c(t)}{\langle O\rangle_{\rm noisy}'-c'(t)}.
    \ee
Here, $\langle O\rangle_{\rm noisy}'$ and $\langle O\rangle$ represent noisy and noiseless expectation values of a ``reference'' observable, measured on the device for the spin-flip interaction strength $\epsilon$ and X-gate rotation angle $\phi$ corresponding to a Clifford point ($\epsilon = 0$ and $\phi = 0$ or $\phi = \pi/2$). 
The four free parameters, i.e., odd-period and even-period values for each offset parameter $c(t)$ and $c'(t)$, are determined via a convex optimization procedure. This procedure fits the proposed ansatz to a classical simulation of the observable on either $2\times2$ or $3\times3$ heavy hexagons. 
These parameters are subsequently used to reconstruct $\langle O\rangle$ on $3\times 7$ heavy hexagons.  
Intuitively, this procedure cancels the multiplicative attenuation factor $f(t)$ by normalizing against Clifford point data, while the additive offsets are learned from classical simulations. 
For further details, as well as extensions to more sophisticated models required to recover the mean square values of two-point correlators and Hamming distances, we refer the reader to Supplementary Information, Sections~\ref{sec:supp_signalrecov_correlations} and ~\ref{sec:supp_signalrecov_hamming}.

\subsection{Classical simulation}
\label{sec:supp_methods_classical}
We employ Matrix Product State (MPS) methods by mapping the decorated hexagonal topology onto a one-dimensional chain with long-range interactions. 
In this approach, the unitary operators in Eq.~\eqref{eq:xxz_layers} are represented as Matrix Product Operators (MPOs). 
We then utilize standard algorithms to contract an MPO with an MPS \cite{paeckel2019time,schollwock2011density}, enabling us to obtain a representation of the time-evolved wave function.
While the bond dimension of the MPS representing the state grows rapidly with each application of $U_F$, the bond dimensions required to express the MPOs for each $U_{\rm XXZ}^{(k)}$ remain relatively modest. 
These bond dimensions primarily depend on how the two-dimensional topology is mapped onto the one-dimensional chain. 
A key limitation of the MPS approach is that the bond dimension of the MPO can grow exponentially with the number of overlapping bonds. 
We outline strategies to minimize this effect in Section~\ref{sec:supp_mat_2dtns} below; see also Ref.~\cite{kim2023evidence}. 
This method is implemented using the TeNPy software package \cite{hauschild2018efficient, hauschild2024tensor}.  

We also utilize classical methods that leverage a tensor network architecture matching the topology of our model \cite{tindall2024efficient}. 
Specifically, we employ an algorithm based on belief propagation \cite{tindall2023gauging} to determine a gauge transformation that maps the two-dimensional network into an approximate canonical form, referred to as the ``Vidal-gauge" in \cite{tindall2023gauging}, inspired by related work in one dimension \cite{vidal2003efficient}. 
Single-site observables are computed by applying a rank-one approximation to the environment of the site; see Ref.~\cite{tindall2023gauging} and Section~\ref{sec:supp_mat_2dtns} for further details. 
Higher-weight observables can also be evaluated using this approach by appending a Clifford circuit to the state and measuring the corresponding single-site operator. 
This procedure reproduces the expectation value of a two-site operator in the original state, albeit at the cost of increasing the entanglement within the system. 
For example, we employ this method to compute two-point correlation functions below. 
These calculations are implemented using the software package \textbf{ITensorNetworks.jl} \cite{iTensorNetworks}, which is built on top of the package \textbf{ITensors.jl} \cite{ITensor}.

\begin{table}[!ht]
    \centering
    \begin{tabular}{|l|c|c|c|c|c|}
    \hline
        Date & $T_1$ ($\mu$s)  & $T_2$ ($\mu$s) & RO ($10^{-2}$)& 1Q ($10^{-4}$) & 2Q ($10^{-3}$) \\ \hline
July 26th, 2024 (Mean) & $110$ & 90 & $2.4$ & $3.6$ & $18.0$ \\ \hline
        July 26th, 2024 (Median) & $100$ & 94 & $1.6$ & $2.9$ & $3.0$ \\ \hline
        July 27th, 2024 (Mean) & $120$ & 93 & $2.4$ & $3.0$ & $18.0$ \\ \hline
        July 27th, 2024 (Median) & $110$ & 91 & $1.7$ & $2.6$ & $3.0$ \\ \hline
        December 3rd, 2024 (Mean) & $130$ & 90 & $2.3$ & $3.0$ & $18.0$ \\ \hline
        December 3rd, 2024 (Median) & $130$ & 84 & $1.7$ & $2.6$ & $2.9$ \\ \hline
        December 11th, 2024 (Mean) & $120$ & 100 & $2.2$ & $3.1$ & $29.0$ \\ \hline
        December 11th, 2024 (Median) & $120$ & 95 & $1.6$ & $2.7$ & $3.0$ \\ \hline
        December 17th, 2024 (Mean) & $140$ & $100$ & $2.1$ & $2.9$ & $18.0$ \\ \hline
        December 17th, 2024 (Median) & $130$ & 97 & $1.7$ & $2.4$ & $2.9$ \\ \hline
        December 18th, 2024 (Mean) & $140$ & $110$ & $2.2$ & $3.7$ & $23.0$ \\ \hline
        December 18th, 2024 (Median) & $140$ & $100$ & $1.6$ & $2.5$ & $3.0$ \\ \hline
    \end{tabular}
    \caption{Device specifications, in particular  relaxation times $T_1$,  dephasing time $T_2$, readout (RO), single-qubit (1Q) errors and two-qubit (2Q) errors rates are reported for the dates when the quantum backend was accessed for data collection.}
\label{tab:device_details}
\end{table}

\subsection{Implementation details}
\label{sec:supp_methods_implementation}
The circuits were designed and parameterized by $c(\phi,\epsilon,s,T)$ for a particular $\phi$, $\epsilon$ point on the phase diagram, where $s$ is the index of the seed used to calculate the disorder in $J$, and $T \in \left[0,T_{max}\right]$ is the number of Floquet repetitions. 
Thus, a total of $T_{max}$ circuits are required to characterize the dynamics of each point on the phase diagram for a given disorder set. The circuits were then transpiled with the preset pass manager optimization level set to 1. 
This level of transpilation includes single-qubit gate optimization and inverse cancellation. 
The layout transpilation step was disabled to ensure a manual mapping of physical to virtual qubits. 
To limit the memory required for the auxiliary classical computing steps for each job after circuit submission, each set of circuits was divided into several circuit splits, for a total of $N_{cs}$ circuit splits per run. 
A predetermined fractional number of $N_{cs}$, along with all circuits for the two Clifford points $\epsilon=0.0$, $\phi \in [0,0.5\pi]$, were submitted using \verb+Session+ in Qiskit for a total of $N_{S}$ sessions. 
All sessions used the \verb+SamplerV2+ Qiskit primitive. 
For all sessions, device-level dynamical decoupling, Pauli gate twirling, and Pauli measurement twirling were disabled. Details of each run are given in Table~\ref{tab:exp_details}.

\begin{table}[!ht]
\centering
\begin{tabular}{|c||c|c|c|c|c|c|c|c|}
\hline
Figure & Geometry & $T_{max}$ & $N_{q}$ & $N_{shots}$ & $N_{S}$ & $N_{cs}$ & Selected Qubits & Start Date of Runs \\
\hline
 Fig.~\ref{fig:fig2} & 2x2 & 50 & 35 & 20000 & 1 & 4 & Fig.~\ref{fig:fez_2x2_3x3_qubits_50cycle} & December 11th, 2024 \\
 Fig.~\ref{fig:fig2} & 3x3 & 50 & 68 & 20000 & 1 & 6 & Fig.~\ref{fig:fez_2x2_3x3_qubits_50cycle} & December 3rd, 2024\\
 Fig.~\ref{fig:fig2} & 3x7 & 50 & 144 & 20000 & 1 & 14  & Fig.~\ref{fig:fez_connectivity_2} & December 3rd, 2024\\
 Fig.~\ref{fig:phase_diagr} & 2x2 & 30 & 35 & 5000 & 33 & 165  & Fig.~\ref{fig:fez_2x2_3x3_qubits} & July 26th, 2024 \\
 Fig.~\ref{fig:phase_diagr} & 3x3 & 30 & 68 & 5000 & 33 & 165  & Fig.~\ref{fig:fez_2x2_3x3_qubits} & July 27th, 2024 \\
 Fig.~\ref{fig:phase_diagr} & 3x7 & 30 & 144 & 5000 & 33 & 165  & Fig.~\ref{fig:fez_connectivity_1} & July 20th, 2024 \\ 
 Fig.~\ref{fig:polarized_state} & 2x2 & 30 & 35 & 5000 & 33 & 165 & Fig.~\ref{fig:fez_2x2_3x3_qubits} & December 18th, 2024 \\
 Fig.~\ref{fig:polarized_state} & 3x7 & 30 & 144 & 5000 & 33 & 165 & Fig.~\ref{fig:fez_connectivity_1} & December 17th, 2024\\
 \hline
\end{tabular}
\caption{Details of each run used for data for a corresponding figure in the main text. $T_{max}$ is the maximum number of Floquet cycles, $N_{q}$ is the number of physical qubits, $N_{shots}$ is the number of shots for each circuit, $N_{S}$ is the total number of sessions for the run, $N_{cs}$ is the total number of circuit splits used for the run, the selected qubits corresponds with the corresponding figure in this Supplementary, and the state date of the runs is for the first session of the run.}
\label{tab:exp_details}
\end{table}

\begin{figure}[!ht]
    \vspace{4cm}
    \centering
    \includegraphics[width=\textwidth]{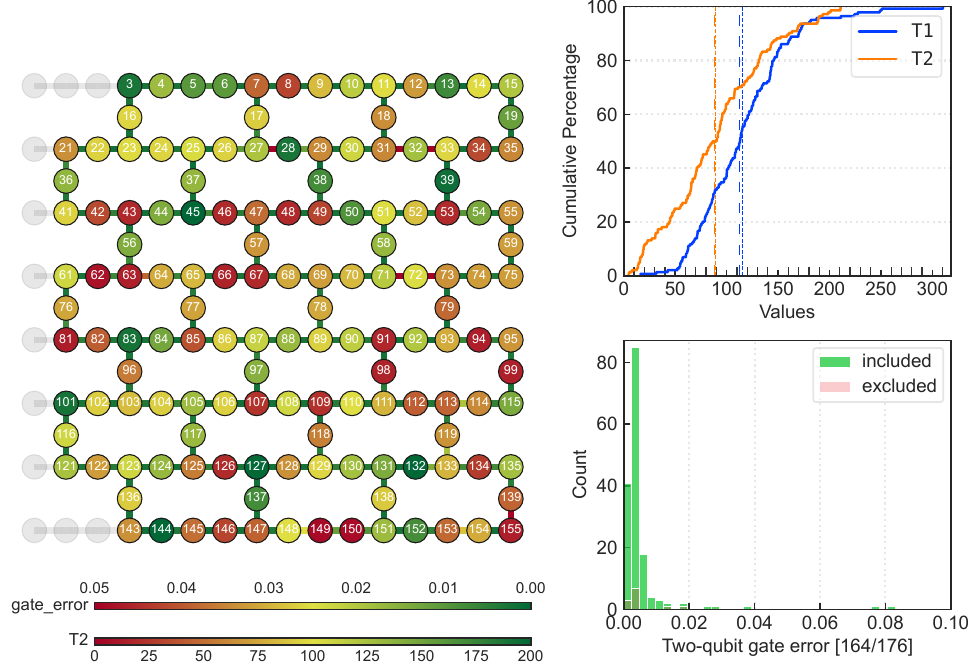}
    \caption{Device layout showing the data qubits for the 3x7 geometry and their specifications for our experiment on $\texttt{ibm\_fez}$ for July 27th, 2024 corresponding to the data shown in Figure \ref{fig:phase_diagr}. Left panel: Data qubits and the corresponding edges are highlighted according to their measured error rates ($T_2$ and two-qubit gate error). Top right panel: Coherence times $T_1$ and $T_2$ are shown using a cumulative distribution plot. Mean and median are highlighted using dotted and dashed lines. Bottom right panel: Two-qubit error rates are shown in a histogram; included and excluded qubits are plotted separately.}  \label{fig:fez_connectivity_1}
    \vspace{4cm}
\end{figure}
\begin{figure}[!ht]
    \vspace{4.6cm}
    \centering
    \includegraphics[width=\textwidth]{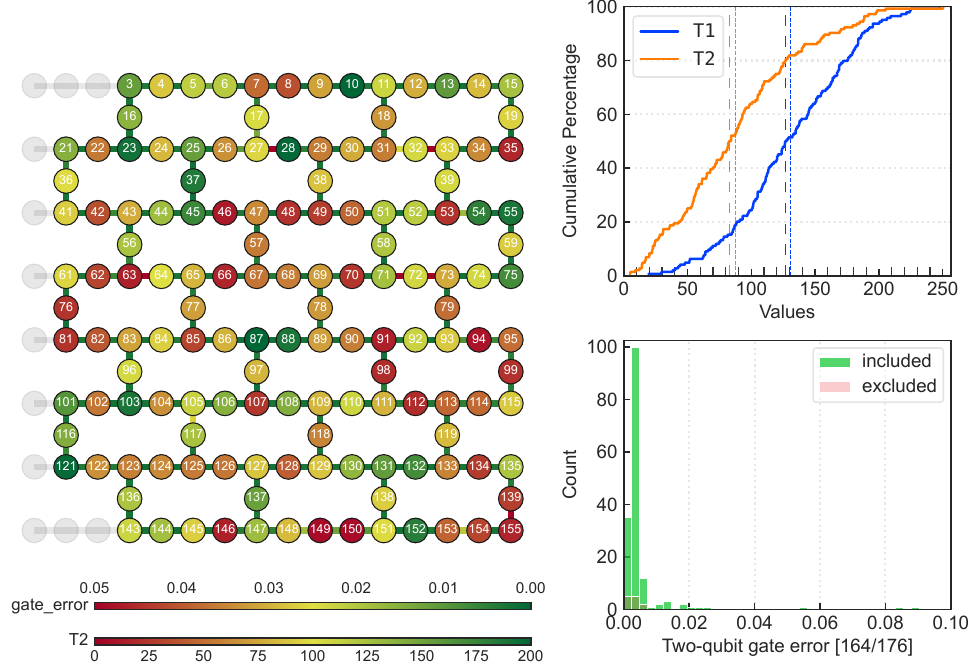}
    \caption{Device layout showing the data qubits for the 3x7 geometry and their specifications for our experiment on $\texttt{ibm\_fez}$ for December 3rd, 2024 corresponding to the data shown in Figure \ref{fig:fig2}. }  \label{fig:fez_connectivity_2}
    \vspace{4.6cm}
\end{figure}

\section{Device details}
\label{sec:supp_device}
For our experiments, we used $\texttt{ibm\_fez}$, a Heron r2 processor. 
Heron r2 processors comprise 156 fixed-frequency transmon qubits with tunable couplers on a heavy-hex lattice layout. 
At the time of usage, IBM Fez could implement 195,000 circuit layer operations per second. 
Of the 156 qubits on Fez, 144 qubits were chosen as our data qubits. 
Figures \ref{fig:fez_connectivity_1} and \ref{fig:fez_connectivity_2} highlight the data qubits such that the vertex of the connectivity graph shows the $T_2$ coherence times for the qubits and the edges show the error rates of two qubit gates. 
Figures \ref{fig:fez_connectivity_1} and \ref{fig:fez_connectivity_2} correspond to the quantum backend's properties on the two data collection days - July 27th 2024 and December 3rd, 2024. The detailed device specification for all the data collection days are shown in Table \ref{tab:device_details}. 
The qubit selection for the MPS labelings of $2 \times 2$, $3 \times 3$, $3 \times 7$ heavy hexagons are shown in Figure \ref{fig:2x2_mps}. 
Let us consider the device specifications for our data qubits for December 3rd 2024. The mean readout error was $2.3 \%$ and the median was $1.7 \%$.  The higher mean indicates an asymmetry with some high readout error qubits skewing the distribution. 
Likewise, the dephasing time $T_2$ had a mean of $90\,\mu \text{s}$ and a median of $84\,\mu \text{s}$. 
In contrast, the relaxation times $T_1$ had a more symmetric distribution with both mean and median times of $130\,\mu \text{s}$. 
As expected, the dephasing times are lower than the relaxation times.  
The single qubit gates had mean error rates of $3.0 \times 10^{-4}$ and median $2.6 \times 10^{-4}$.
A similar qubit-wise asymmetry as discussed above was seen here as well. However, the difference between the mean and the median is not substantial. 
Lastly, the two-qubit gate error had a mean of $1.8 \times 10^{-2}$ and median $2.9 \times 10^{-3}$.

\section{Classical simulation}
\label{sec:supp_classicalsim}
We used two classes of tensor network-based algorithms for our classical simulations.
The first class utilizes Matrix Product States (MPS) along with established algorithms for updating, truncating, and contracting the MPS.
The second class involves techniques based on two-dimensional tensor network states, which are designed to align with the topology of the decorated hexagonal lattice. In this case, we adopt algorithms inspired by Belief Propagation \cite{tindall2023gauging, tindall2024efficient} to update the two-dimensional tensor network. Further details on both classes of methods are provided in Sections \ref{sec:supp_mat_mps} and \ref{sec:supp_mat_2dtns} below.

The MPS representation provides one of the most common and well-understood techniques for classical simulation, primarily applied to one-dimensional systems. The success of these methods ultimately boils down to two key properties of MPS. First, MPS contraction is efficient; more precisely, the computational cost of contraction scales as $\chi^3$, where $\chi$ is the maximum bond dimension of the network.  
Second, tensor network states generally possess a redundant degree of freedom known as gauge freedom, meaning that for a given quantum state, the choice of tensors forming the network representation of the state is not unique.  
By applying a “gauge transformation,” one can transform one tensor network into another with more favorable properties without affecting the underlying physical state.  
If the tensor network under consideration is an MPS, it is always possible to find a gauge transformation that converts the MPS into a ``canonical form" such that each bond of the network defines an orthogonal basis.  
This allows for the truncation of the bond dimension at any bond in the network in an optimal way and with a controlled truncation error, see Ref.~\cite{schollwock2011density} for a review. The MPS representation can also be extended to models in two or higher dimensions by `unrolling' them onto a single dimension. However, this approach introduces long-range interactions within the model, rendering standard algorithms, such as TEBD \cite{vidal2004efficient}, inapplicable. To deal with these long range interactions, one can use approaches such as TDVP \cite{haegeman2011time, haegeman2016unifying}, the $W^{I, II}$ method \cite{zaletel2015time}, or the direct construction of the Matrix Product Operator (MPO) representing the time evolution operator—see Section~\ref{sec:supp_mat_mps} below.  

One can avoid the introduction of long-range interactions by using a two-dimensional tensor network state that matches the topology of the Hamiltonian under consideration. However, there are two primary obstacles facing two-dimensional tensor network states. First, the contractions required to calculate meaningful quantities, such as expectation values of observable quantities, scale exponentially with the size of the smallest spatial dimension \cite{schuch2007computational}. Second, while methods exist to gauge-transform a general two-dimensional tensor network into a canonical form analogous to that of MPS \cite{evenbly2018gauge, haghshenas2019conversion, tindall2023gauging}, these higher-dimensional canonical forms do not allow for optimal or even well-controlled truncations of the bond dimension, see Section~\ref{sec:supp_mat_2dtns} below. 

In Figures~\ref{fig:3x7_pol} and \ref{fig:3x7_corr} we compare the results for the single site polarizations and nearest neighbour two point correlators obtained by the two classical simulation methods, i.e. Matrix Product State simulations and two-dimensional tensor network state simulations with Belief Propagation (BP). In both figures, we compare the results for a number of different bond dimensions. Note that in all the plots there is excellent agreement between the methods at early to intermediate times. However, the results from the two methods begin to diverge from each other at later times. Furthermore, we observe that, at these later times, the MPS based simulations often struggle to converge with increasing bond dimension, whereas this is not the case for the two-dimensional tensor network state simulations which were updated and contracted with Belief Propagation.

\subsection{Matrix Product State simulations}
\label{sec:supp_mat_mps}

\begin{figure}[t!]
    \centering
    \includegraphics[width=.46\textwidth]{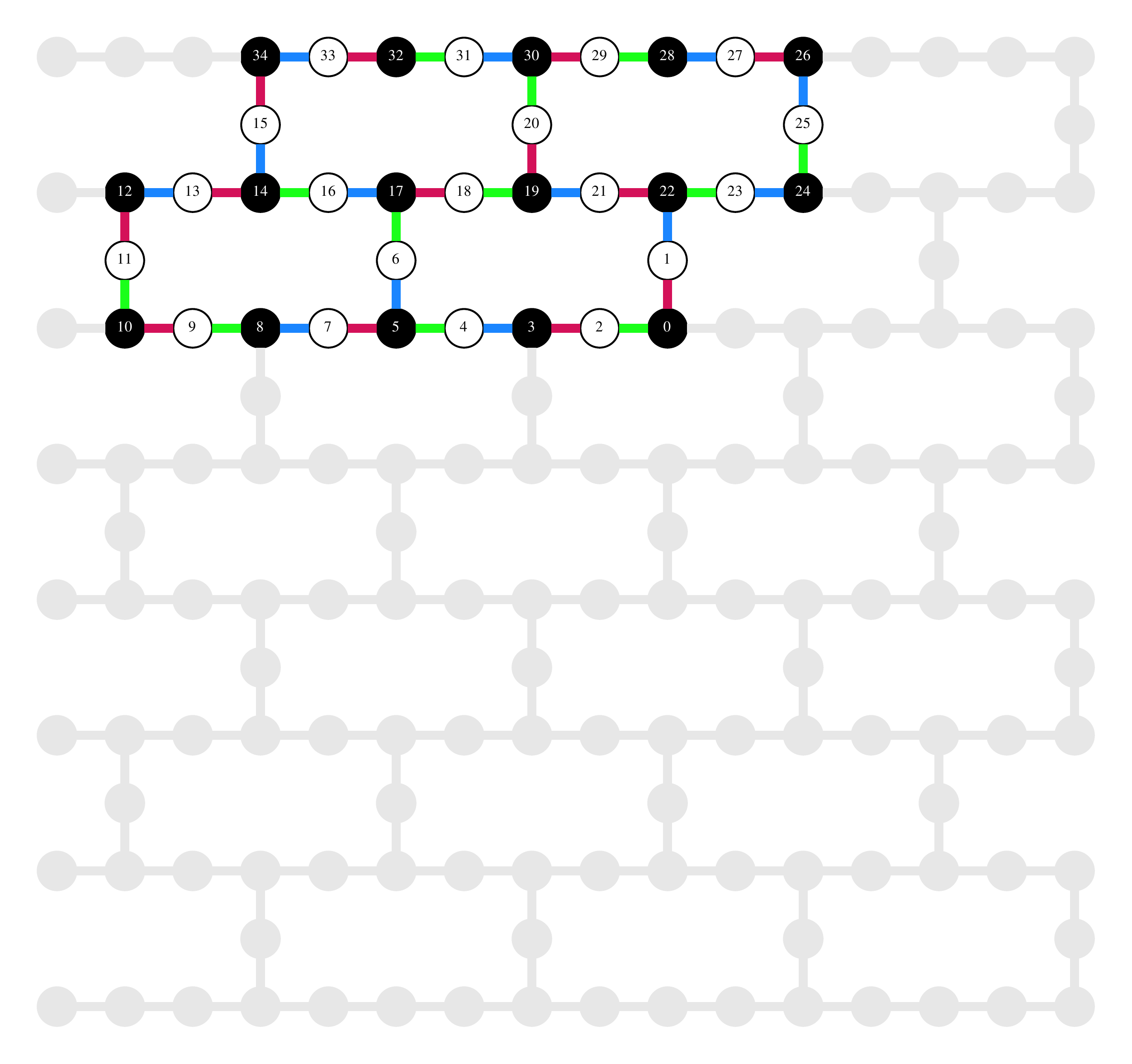}
    \includegraphics[width=.46\textwidth]{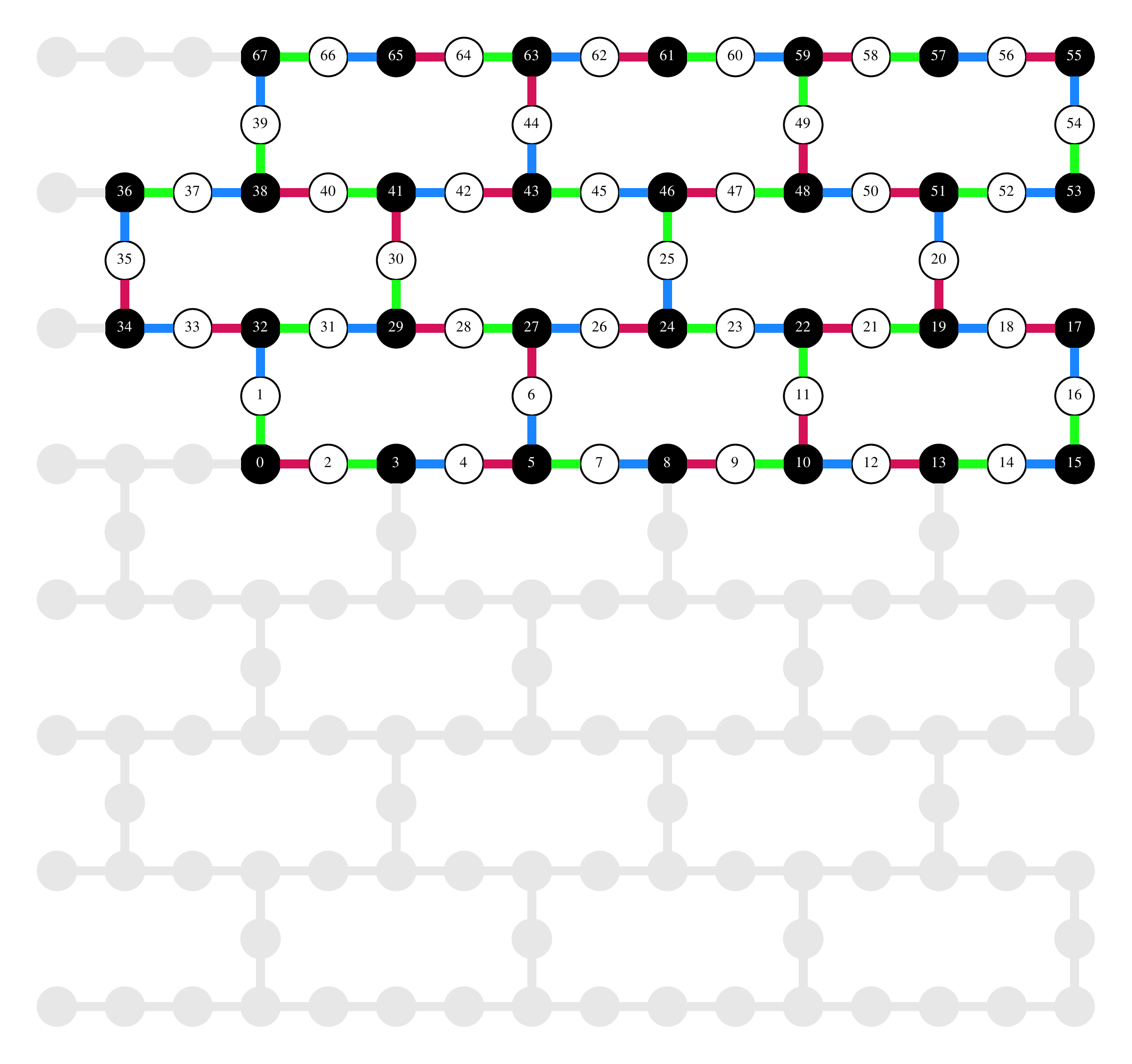}
    \includegraphics[width=.46\textwidth]{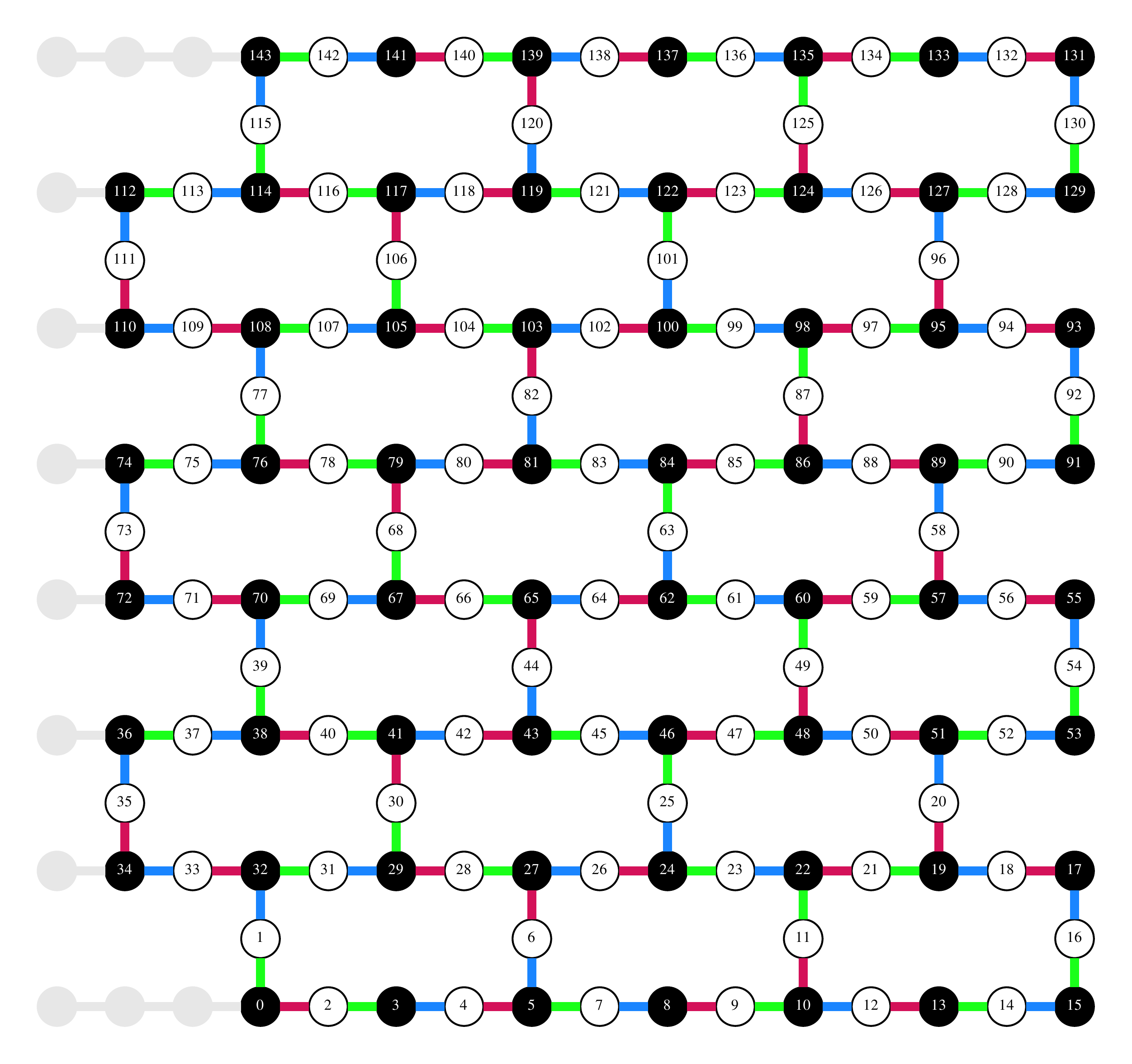}
    \caption{The MPS labeling of the system with $2\times2$, $3\times3$ and $3\times7$ heavy hexagons.}  \label{fig:2x2_mps}
\end{figure}

To apply MPS-based algorithms to two-dimensional models, we first `unroll' the two-dimensional model onto one dimension, i.e. assign an integer value between $0$ and $N-1$ to each of the $N$ sites in the two-dimensional model. This assignment is illustrated in Figure \ref{fig:2x2_mps}.  The unrolling procedure comes at the expense of introducing long-range interactions.  
For example, the qubits with the MPS labels $1$ and $22$ in the top left panel of Figure \ref{fig:2x2_mps} are nearest neighbors on the two-dimensional decorated hexagonal graph but are well separated on the MPS chain.  
A gate that acts on nearest neighbors in the heavy-hex model thus becomes a long-range gate when applied to the MPS representing the quantum state. There is no unique way to unroll the two-dimensional graph onto one dimension, and the precise method used will affect the amount of entanglement (and hence the required bond dimension for a given precision) in the description of the state.  
Here, we use a labeling scheme similar to the one previously used in Ref.~\cite{tindall2024efficient}.  

We implement the long-range gates by constructing the Matrix Product Operator (MPO) for each layer $U_F^{(k)}$ in $U_F$, see Eq. (\ref{eq:UF_def}) in the main text.  
We can represent each two-qubit operator $U_{ij}$ in Eq. (\ref{eq:U_ij_def}) as an MPO with bond dimension $\chi = 4$.  
The bond dimension of each $U_F^{(k)}$ is determined by the number of overlapping bonds after unrolling onto one dimension.  For example, a maximum bond dimension of $64$ is required to store the MPO of any of the individual layers $U_F^{(j)}$ in the system with $2\times2$ heavy hexagons in Figure \ref{fig:2x2_mps}.   
In all our simulations, the initial quantum state is a product state, which can be represented as a trivial MPS with bond dimension $1$.  The MPOs representing $U_F^{(1)}$, $U_F^{(2)}$, and $U_F^{(3)}$ are then individually contracted with the MPS representing the quantum state, such that we never need to store the MPO representing $U_F$ itself.  
We perform the MPO-MPS contractions using a standard SVD-based algorithm \cite{schollwock2011density}, in which the MPS is truncated to its maximum allowed value $\chi_{max}$. The MPS simulations with different values of $\chi_{max}$ are compared in Figures \ref{fig:3x7_pol} and \ref{fig:3x7_corr} for one particular point in the phase diagram.

\subsection{Two-dimensional Tensor Network State simulations}
\label{sec:supp_mat_2dtns}
\begin{figure}[t!]
    \centering
    \includegraphics[scale=0.75]{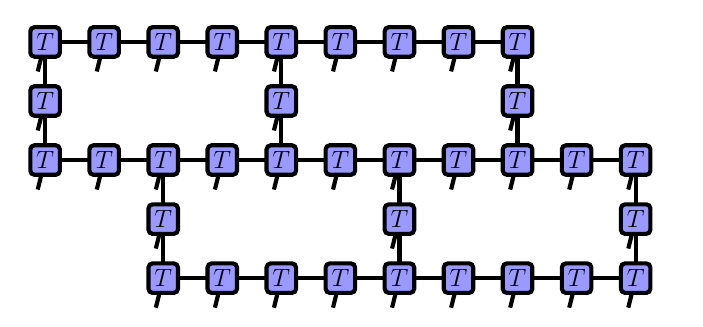}
    \caption{The two-dimensional tensor network state with $2\times2$ heavy hexagons.}  \label{fig:2x2_tn}
\end{figure}

\begin{figure}[t!]
    \centering
\includegraphics[scale=0.75]{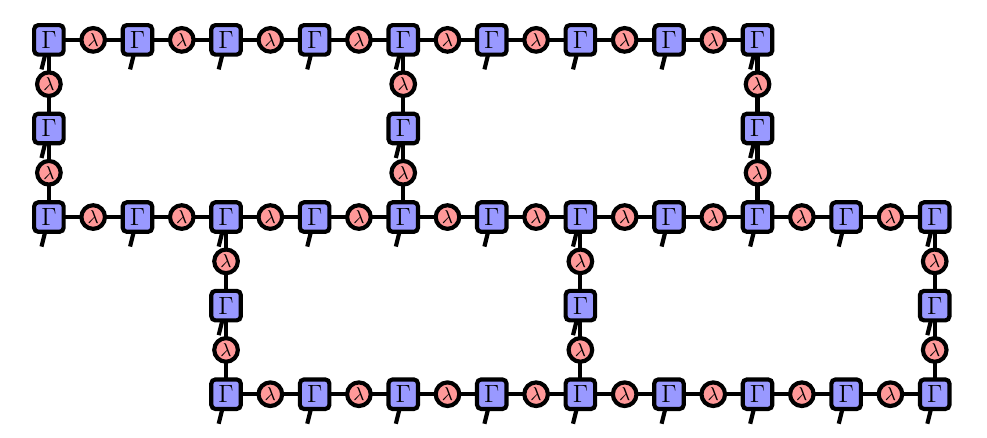}
    \caption{The quasi-canonical form of the the two-dimensional tensor network state with $2\times2$ heavy hexagons. Each site has a vertex tensor $\Gamma$ and each edge has a bond tensor $\lambda$. We use a belief propagation algorithm to ensure that the $\Gamma$ and $\lambda$ tensors approximately satisfy the constraint shown in Figure \ref{fig:vidal_constraint} at each site.}  \label{fig:2x2_tn_canon_form}
\end{figure}

\begin{figure}[t!]
    \centering
\includegraphics[scale=0.75]{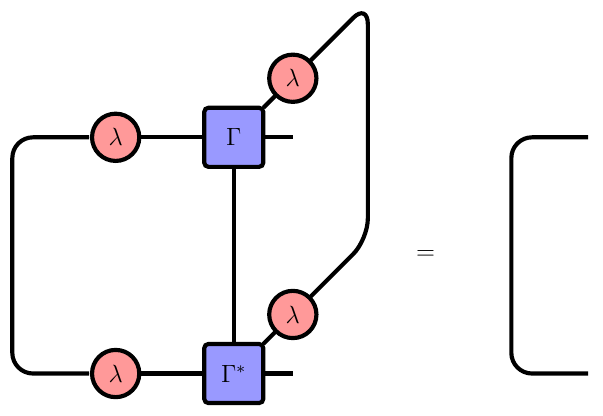}
    \caption{In the Vidal gauge, the vertex and bond tensors at each site satisfy the constraint defined by the contraction diagram shown above. This constraint can be stated as follows: at any given site, if one takes the vertex tensor $\Gamma$ and contracts it with all but one of its bond tensors $\lambda$, the resulting tensor should be an isometry - i.e. contracting with its complex conjugate (LHS above) should result in the identity tensor (RHS above). The belief propagation algorithm applies a gauge transformation to the tensor network such that the equality shown in this contraction diagram is (approximately) satisfied at each site of the network. See \cite{tindall2023gauging} for more details.}  \label{fig:vidal_constraint}
\end{figure}

We consider a two-dimensional tensor network state that matches the decorated hexagonal topology.  
In Figure \ref{fig:2x2_tn}, we show the structure of this tensor network for the case of $2\times2$ heavy hexagons.  
Each tensor has one physical leg with two indices (corresponding to the $\ket{0}$ and $\ket{1}$ states of the qubit at each site) and either two or three virtual legs.  
Note that tensors on different sites are, in general, not equal to each other. The two-qubit operators $U_{ij}$ in Eq.~\eqref{eq:U_ij_def} in the main text can be implemented as gates that update pairs of nearest-neighbor tensors in the two-dimensional network.  
The application of a gate increases the bond dimension of the state, which we then truncate to the maximum bond dimension $\chi_{max}$.  

The error incurred by truncation depends heavily on the gauge of the tensor network, which we now discuss.  For any given quantum state $\ket{\psi}$ described by a tensor network, the choice of tensors $T$ in Figure \ref{fig:2x2_tn} is not unique.  For example, one can redefine each tensor $T$ via the transformation $T \rightarrow X^{-1}TX$ for some matrix $X$.  
If the same $X$ is used for each tensor $T$, then the state $\ket{\psi}$ described by the network remains unchanged under this gauge transformation, see Ref.~\cite{hauschild2018efficient} for a more extensive discussion of this point.  

For MPS, it is well known how to construct the gauge for the network such that the error incurred by truncating the bond dimension is optimal, as defined by the fidelity between the state before and after truncation \cite{schollwock2011density}. This is known as the canonical form or the ``Vidal gauge." There is no simple way to construct a canonical form that can be truncated optimally for more general tensor network states, such as the one in Figure \ref{fig:2x2_tn}.  
Instead, one can construct a quasi-canonical form of a two-dimensional tensor network with properties similar to its one-dimensional counterpart, such that truncating the bond dimension produces errors low enough to be practical, even if they are not optimal. In this gauge, there are ``vertex" tensors and ``bond" tensors, referred to as $\Gamma$ and $\lambda$, respectively, as shown in Figure \ref{fig:2x2_tn_canon_form}.  
The vertex tensors have one physical leg and either two or three virtual legs, while the bond tensors have two virtual legs and can be represented as square matrices.  
The tensor network is said to be in the Vidal gauge if all tensors satisfy a particular constraint, i.e. that the combination of the vertex and bond tensors at each site forms an isometry, see Figure \ref{fig:vidal_constraint} for more details.  

When truncating the bond dimension of the network, the truncation error is reduced when this truncation is performed while in the Vidal gauge. 
However, each application of a gate $U_{ij}$ destroys the quasi-canonical form - the network thus has to be repeatedly ``regauged" throughout the simulation. We use a belief propagation algorithm to put the tensor network back into its quasi-canonical form, i.e., so that the constraint in Figure \ref{fig:vidal_constraint} is (approximately) satisfied at each site. 
As discussed in \cite{tindall2023gauging, tindall2024efficient}, regauging after each application of a gate is superfluous and impractically expensive. 
We thus only regauge the network with belief propagation after each Trotter step. 

Once we perform the updates to obtain a representation of the time-evolved quantum state, we need to extract meaningful quantities, such as the expectation values of observable operators. 
Doing so exactly would require the contraction of the full two-dimensional network—a notoriously difficult computational task \cite{schuch2007computational}. 
However, we can calculate one-site observables by approximating the environment of the site as a separable product given by the $\lambda$ matrices attached to the site in question. 
As discussed in detail in \cite{tindall2023gauging}, when there are loops in the network—as is the case here—this approximation may not hold. 
Higher-weight observables can be calculated by applying a Clifford circuit $U_c$ to the network such that the expectation value of a single-site observable, after the application of the Clifford circuit, is equal to the expectation value of the higher-weight observable in the original state \cite{tindall2024efficient}. 
More precisely, if we want to calculate the expectation value of, e.g., the two-point correlator $Z_{i}Z_{j}$, we find the Clifford circuit $U_c$ such that $Z_{i}Z_{j} = U_c^{\dagger}Z_{i}U_c$ and hence $\bra{\psi}Z_iZ_j\ket{\psi} = \bra{\psi}U_c^{\dagger}Z_{i}U_{c}\ket{\psi}$. 
We use this method to obtain the BP results in Figure \ref{fig:3x7_corr}. 
This technique comes at the expense of increasing the number of gates applied to the network, thereby increasing the amount of entanglement in the state and the number of truncations that must be applied.

\begin{figure}[t!]
    \centering
    \includegraphics[width=.46\textwidth]{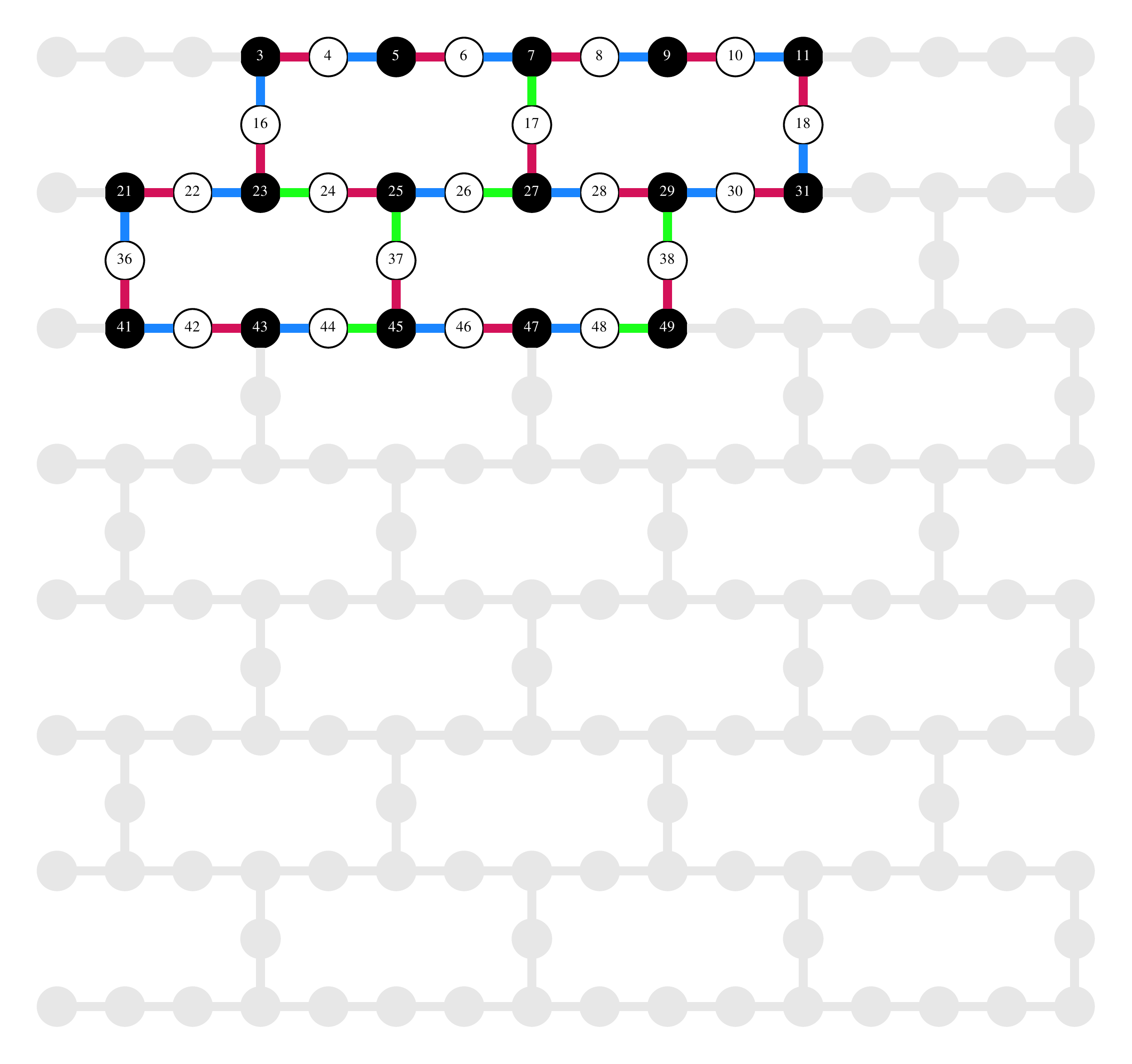}
    \includegraphics[width=.46\textwidth]{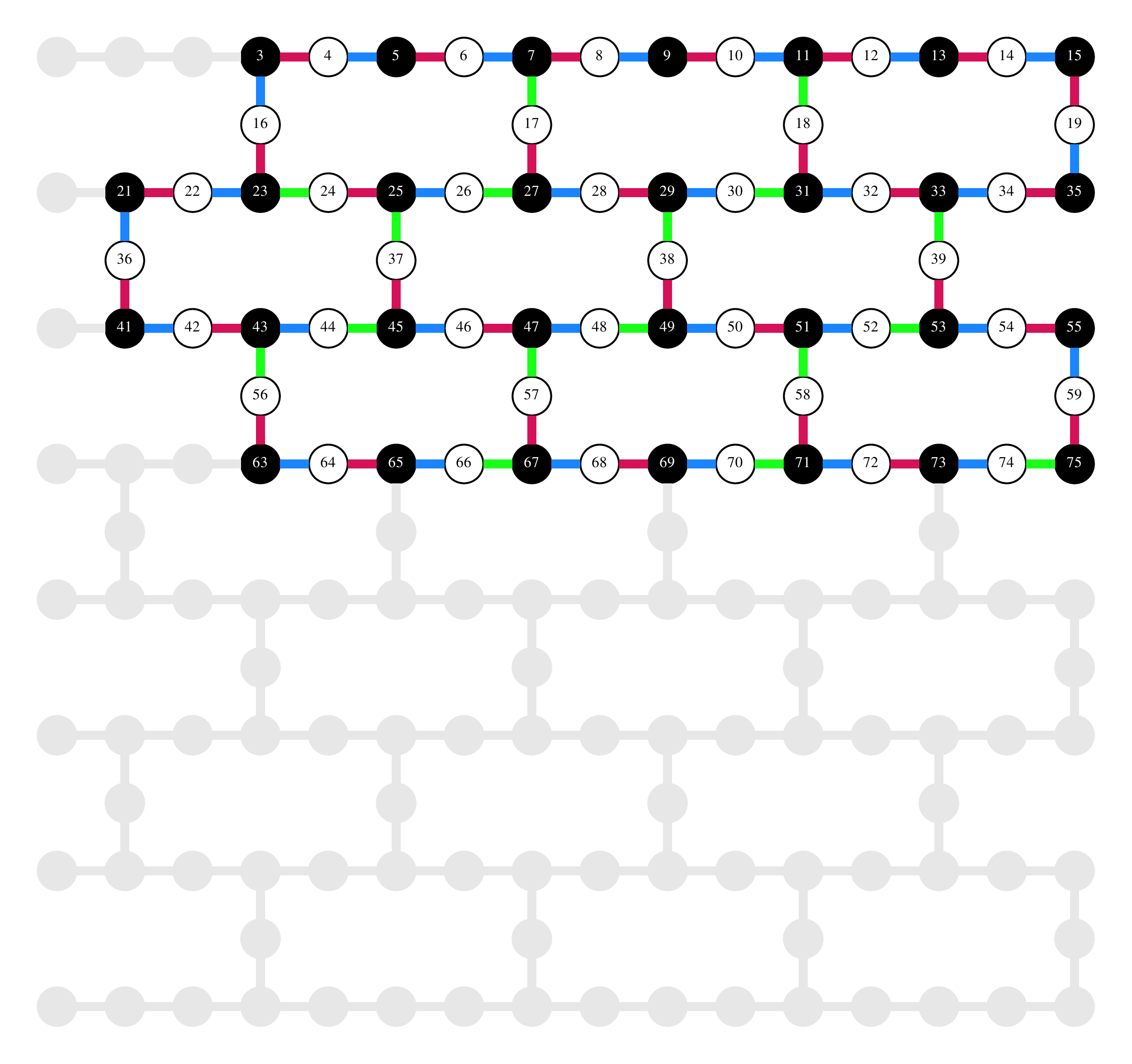}
    \caption{Selected device qubits for $2\times2$ and $3\times3$ geometries corresponding to Figure \ref{fig:phase_diagr}. }  
    \label{fig:fez_2x2_3x3_qubits}
\end{figure}

\begin{figure}[t!]
    \centering
    \includegraphics[width=.46\textwidth]{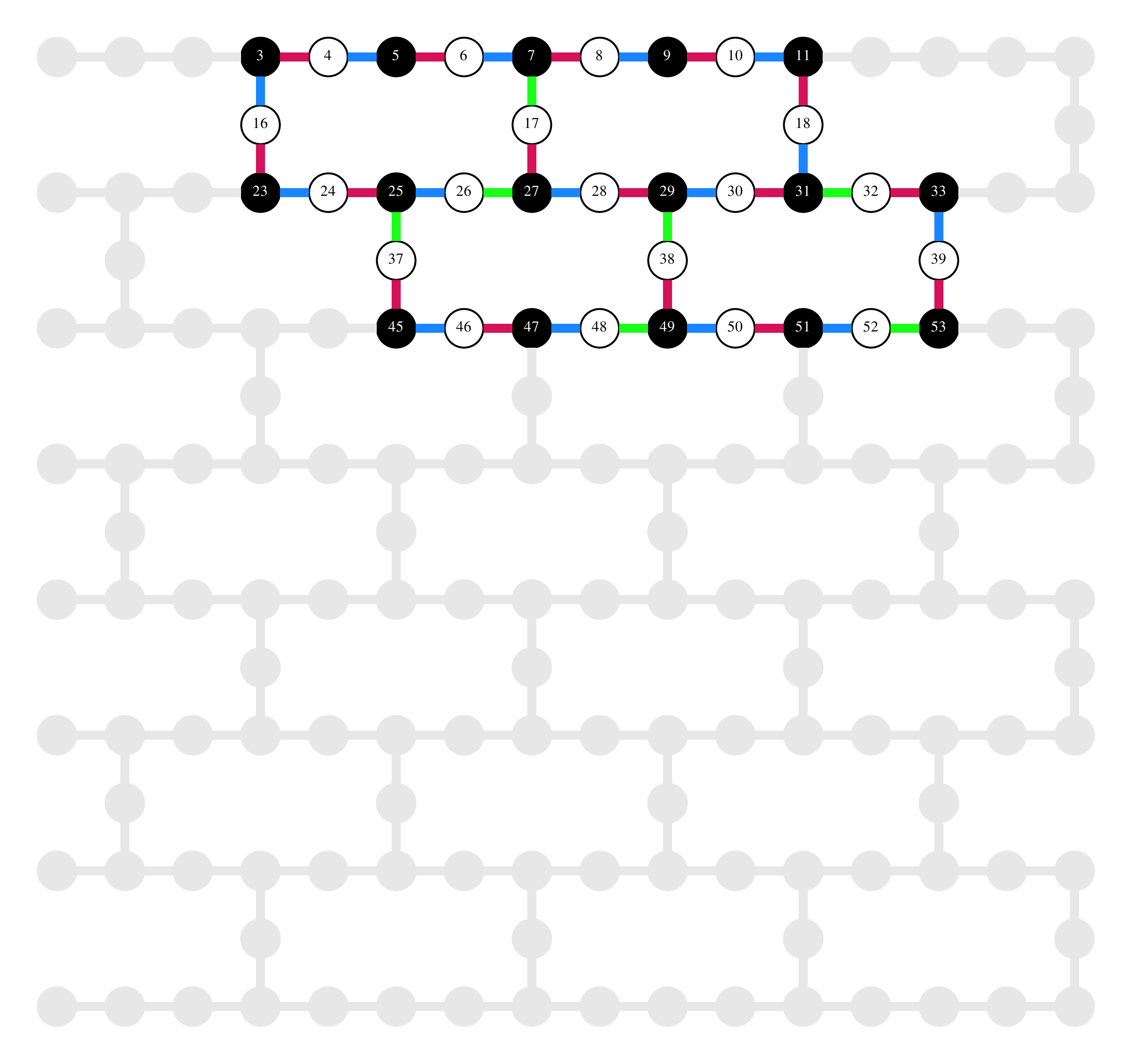}
    \includegraphics[width=.46\textwidth]{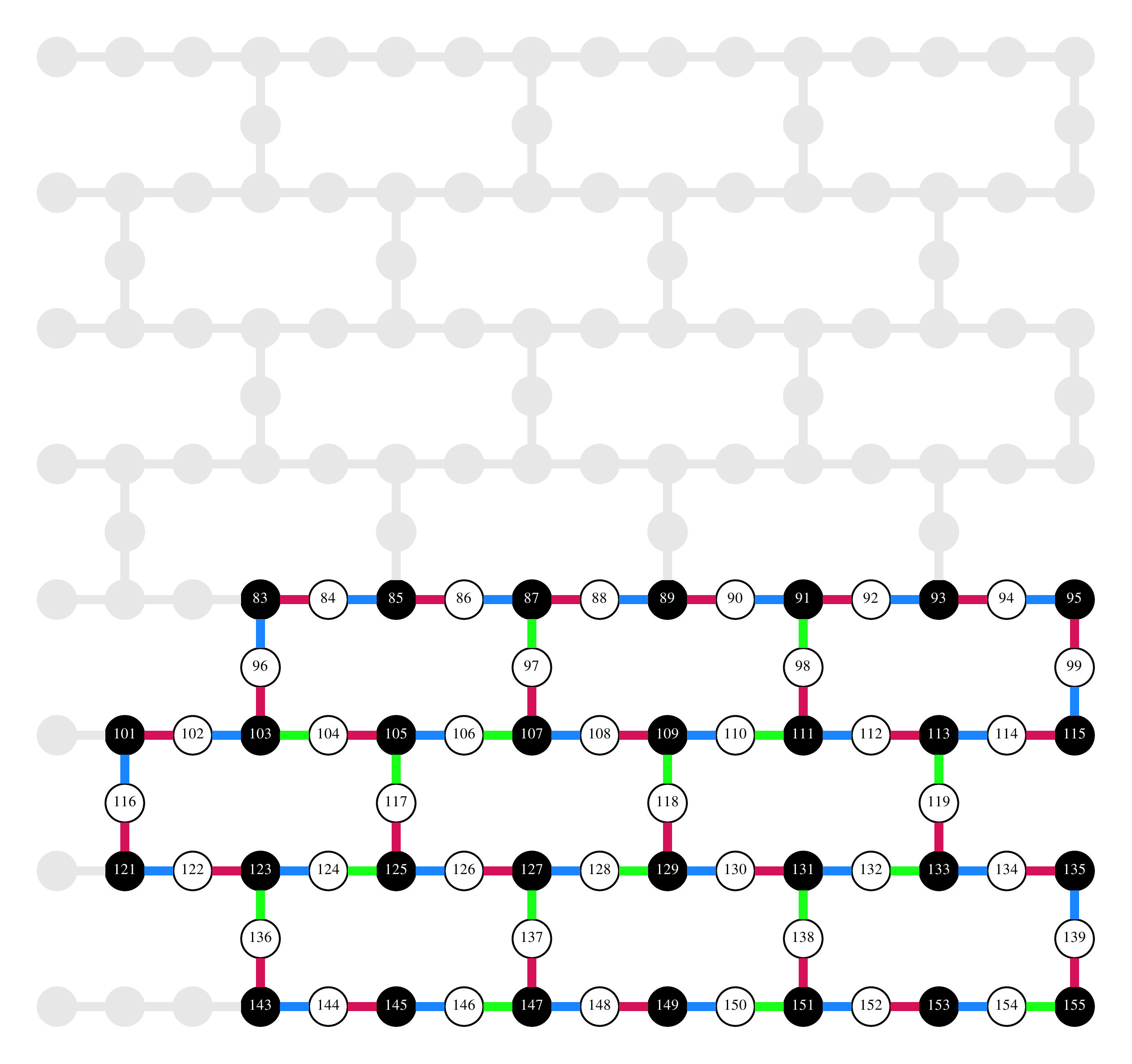}
    \caption{Selected device qubits for $2\times2$ and $3\times3$ geometries corresponding to Figure \ref{fig:fig2}. }  \label{fig:fez_2x2_3x3_qubits_50cycle}
\end{figure}

\begin{figure}[h!]
    \centering
    \includegraphics[scale=0.425]{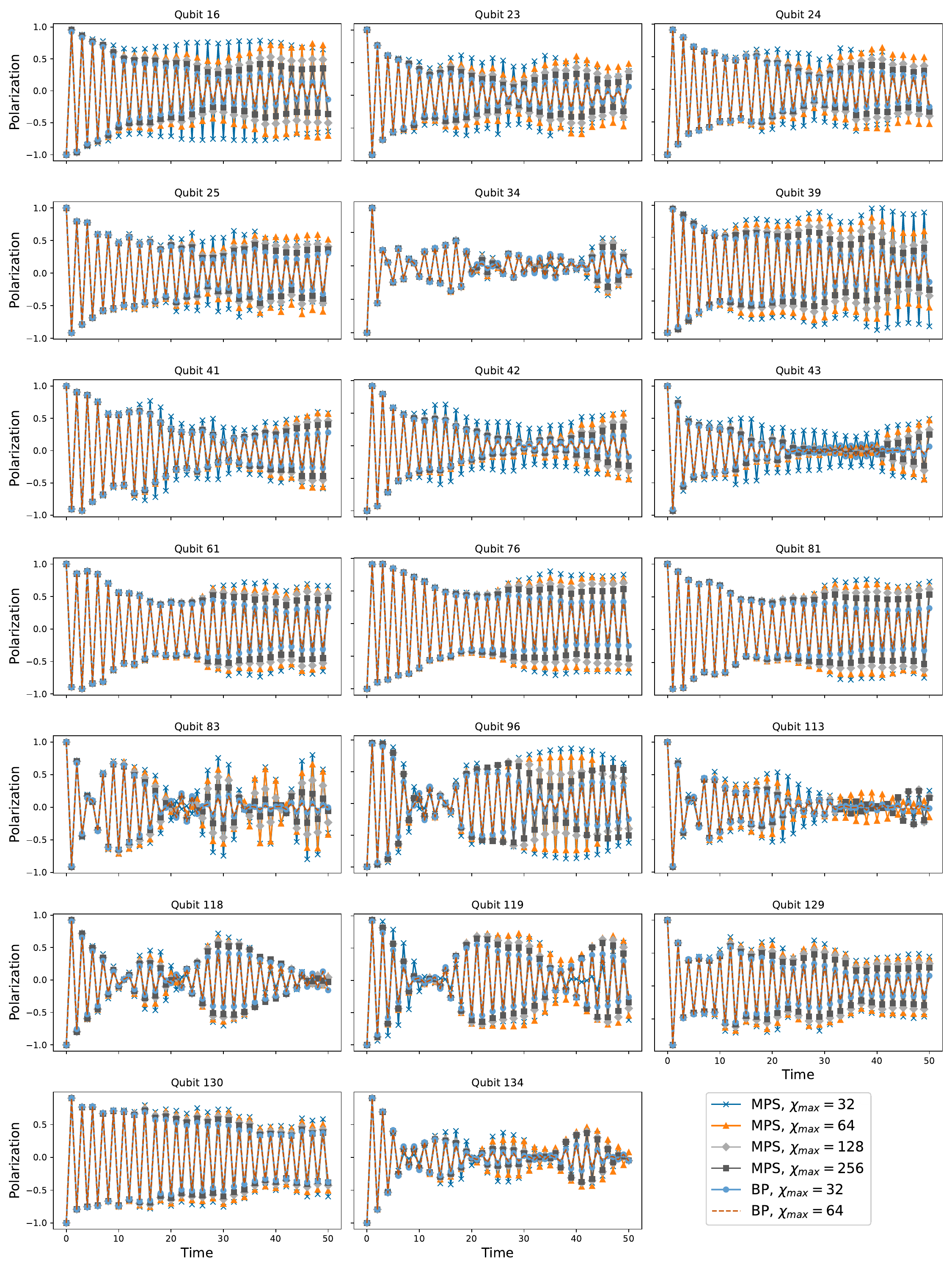}
    \caption{Classical simulation results for selected site polarizations in the 3x7 geometry for $\epsilon = 0.05$ and $\phi = 0.45\pi$. The simulations based on two-dimensional tensor network states with Belief Propagation (BP) converge with increasing bond dimension, whereas the MPS-based simulations don't always converge with the bond dimensions used here, particularly at later times.}  \label{fig:3x7_pol}
\end{figure}

\begin{figure}[h!]
    \centering
    \includegraphics[scale=0.425]{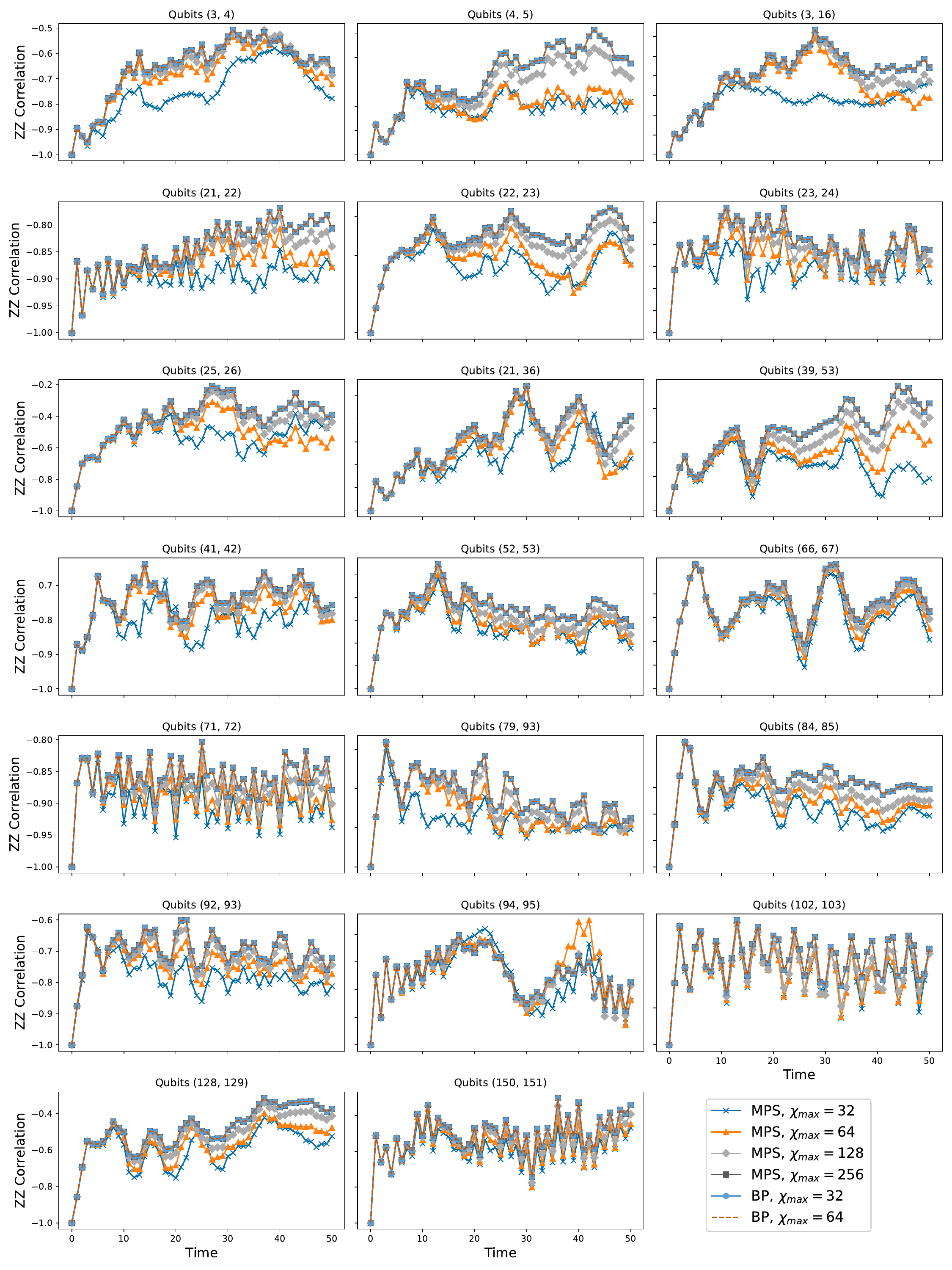}
    \caption{Classical simulation results for $ZZ$-correlations between a selection of nearest-neighbor sites in the 3x7 geometry for $\epsilon = 0.05$ and $\phi = 0.45\pi$. Similarly to the polarizations in Figure \ref{fig:3x7_pol}, the simulations based on two-dimensional tensor network states with Belief Propagation (BP) converge with increasing bond dimension, whereas the MPS-based simulations don't always converge with the bond dimensions used here, particularly at later times.}  \label{fig:3x7_corr}
\end{figure}

\section{Signal recovery from noisy observables}
\label{sec:supp_signalrecov}

\subsection{Spin ordering}
\label{sec:supp_signalrecov_spinorder}
We focus on the expectation values $s_i(t)$ of the Pauli-$Z$ operators for spin $i$ at discrete time $t$, along with the ideal spin ordering parameter $\Delta(t)$, defined as  
\be\label{eqs:delta_dfn}
s_i(t) := \langle Z_i(t)\rangle \equiv \langle\psi_t|Z_i|\psi_t\rangle, \qquad \Delta(t) = \frac{1}{N}\sum_{i=1}^N s_i(0)s_i(t),
\ee
where $ |\psi_t\rangle $ represents the state vector of the noiseless evolution, and $N$ denotes the number of qubits. In the experiment, however, we can only access the value $ \tilde{s}_i(t)$, which incorporates the effects of noise, as well as initialization and measurement errors.  As a result, the experimentally accessible order parameter reflects these noise contributions. Consequently, the measured order parameter takes the form  
\be\label{eq:noisy_Delta}
\Delta_{\rm noisy}(t) := \frac 1N\sum_{i=1}^N s_i(0)\tilde s_i(t),
\ee
where $s_i(0)$ corresponds to the initial spin projections.

While predicting individual $s_i(t)$ from their noisy counterparts $\tilde{s}_i(t)$ remains a challenging problem, several methods currently enable partial recovery of noiseless values, provided multiple copies of the noisy circuit and its modifications are available.  
These methods include twirling \cite{Wallman2017Twirling}, zero-noise extrapolation~\cite{Temme2017PECandZNE,Li2017ZNE}, and probabilistic error amplification and error cancellation~\cite{van2023probabilistic}.  
However, even with this full arsenal of techniques, achieving reliable results beyond $T_1$/$T_2$ timescales remains difficult.  In what follows, we explore an alternative, physics-inspired approach that facilitates the extraction of collective observables — specifically, the average of multiple $s_i(t)$ — for circuit depths up to 450.  
This method relies on the following simplified noise model,  
    \be
    \tilde{s}_i(t) = f_{i}(t)s_i(t) + \delta_{i}(\epsilon,\phi,t),
    \ee
which implies that the exact expectation value $s_i$ is related to the experimentally measured $\tilde{s}_i$ through a linear model with an unknown attenuation coefficient $f_i$ and an additive bias term $\delta_i$. Here, $f_i(t)$ accounts for the unbiased, depolarizing component of the noise, while $\delta_i(\epsilon,\phi,t)$ represents the bias arising from relaxation processes.  
This model effectively approximates sufficiently dense circuits as converting local noise into global white noise.  

In what follows, we do not assume a specific parametric form for $f_i(t)$ (e.g., an exponential), but we neglect its dependence on the parameters $\epsilon$ and $\phi$.  
In this assumptions, 
    \be
    \begin{split}
    \Delta_{\rm noisy}(\epsilon,\phi,t) &= \frac 1N\sum_{i=1}^N s_i(0)\tilde s_i(t)=\frac 1N \sum_{k=1}^N f_{i}(t) s_i(0)s_i(t) + \frac 1N\sum_{k=1}^N s_i(0)\delta_{i}(\epsilon,\phi,t)\\ 
    &= \frac 1N\sum_{i=1}^N (f_{i}(t)-f(t)) s_i(0)s_i(t) +f(t) \frac 1N\sum_{i=1}^N  s_i(0)s_i(t) + \frac 1N\sum_{i=1}^N s_i(0)\delta_{i}(\epsilon,\phi,t)\\
    &=f(t)\Delta(\epsilon,\phi,t) + \delta(\epsilon,\phi,t),\quad f(t) :=  \frac{1}{N}\sum_{i=1}^N f_{i}(t),\quad  \delta(\epsilon,\phi, t) := \frac{1}{N}\sum_{k=1}^N s_i(0)\delta_{i}(\epsilon,\phi,t)+\eta(t)
    \end{split}
    \ee
where we use the notation $\eta(t)=\frac 1N\sum_{i=1}^N (f_{i}(t)-f(t)) s_i(0)s_i(t) $. 
Next, we assume that the signal attenuation function $f_i(t)$ is statistically independent of the noiseless spin values. 
This approximation means that we can introduce two random variables, $\Xi$ and $\Theta$, such that $f_{1}(t)-f(t),\dots, f_N(t)-f(t)$ are observed realizations of $\Xi$, and $s_1(0)s_1(t),\dots, s_N(0)s_N(t)$ -- are realizations of $\Theta$. 
Since $\Xi$ and $\Theta$ are independent, and by the law of large numbers, $\lim_{N\to\infty}\eta(t)=\mathbb E(\Xi) \mathbb E(\Theta)=0$ since $\mathbb E(\Xi)=0$. 
Moreover, using central limit theorem it is not hard to see that $\eta(t)=O\left(\frac{1}{\sqrt{N}}\right)$ for large $N$ and can be neglected.   
As the result, we get the model,
    \be\label{eq:global_noise_polrz}
    \Delta_{\rm noisy}(\epsilon,\phi,t) = f(t)\Delta(\epsilon,\phi,t) + \delta(\epsilon,\phi,t),
    \ee
where $f(t)$ is the overall decay factor and $\delta(\epsilon,\phi,t)$ is the offset. 
It originates from the non-equivalence of logical zero and one states that are usually represented by the lowest and first excited level of the superconducting circuit.  
Since we study the dynamics that breaks the discrete time translation symmetry, we assume that $\delta(\epsilon,\phi,t) = \delta (\epsilon,\phi,t+2T)$. 
This is equivalent to the statement that it takes the form,
    \be
    \delta(\epsilon,\phi,t) = 
    \begin{cases}
    \delta_0(\epsilon,\phi) \quad \text{if} \quad t \in 2\mathbb Z\\
    \delta_1(\epsilon,\phi) \quad \text{if} \quad t \in 2\mathbb Z+1\\
    \end{cases}.
    \ee
Then, using the approximations in Eqs.~\eqref{eq:global_noise_polrz} for two points ($
\epsilon, \phi$) and $(0,\phi_0)$, we arrive at the expression,
    \be\label{eq:polrz_recovered}
    \hat\Delta(\epsilon,\phi,t) = \Delta(0,\phi_0,t) \frac{\Delta_{\rm noisy}(\epsilon,\phi,t) - \delta(\epsilon,\phi,t)}{\Delta_{\rm noisy}(0,\phi_0,t) - \delta(0,\phi_0,t)}.
    \ee
It is convenient to choose $\phi_0$ corresponding to the closest Clifford point. 
This means we take $\phi_0 = 0$ for any $\phi \leq \pi/4$ and $\phi_0 = \pi/2$ for all $\phi > \pi/4$.

The expression in Eq.~\eqref{eq:polrz_recovered} can be used to recover the noiseless value, given that the parameters $\delta_{0,1}(0,\phi_0)$ and $\delta_{0,1}(0,\phi_0)$ are known. 
Unfortunately, these values depend on the noise model and the measurement noise of the device. 
We may assume, however, that these values weakly depend on the system size as the number of qubits increases due to self averaging. 
This provides an opportunity to learn these parameters from smaller system sizes, where classical algorithms achieve sufficiently high precision, and then extrapolate to larger two-dimensional systems with 100+ qubits, where these algorithms become unreliable. 
In what follows we learn offsets $\vec{\delta}=(\delta_0(\epsilon,\phi),\delta_1(\epsilon,\phi),\delta_0(0,\phi_0),\delta_1(0,\phi_0))^\top$ by simulating $\Delta(\epsilon,\phi,t)$ classically, $\Delta_{\mathrm{sim}}(\epsilon,\phi,t)$ and then solving the following convex optimization problem,
    \be\label{eqs:optim_J}
    \min_{\vec{\delta}} J(\vec{\delta})=\sum_{t=1}^T (\Delta_{\mathrm{sim}}(\epsilon,\phi,t)-\hat\Delta(\epsilon,\phi,t))^2+q \|\vec{\delta}\|_2^2,
    \ee
where $q>0$ is a regularization parameter chosen to pick offsets uniquely, with minimal euclidean norm $\|\cdot\|_2^2$.

\begin{figure}[t!]
    \centering
\includegraphics[scale=0.83
]{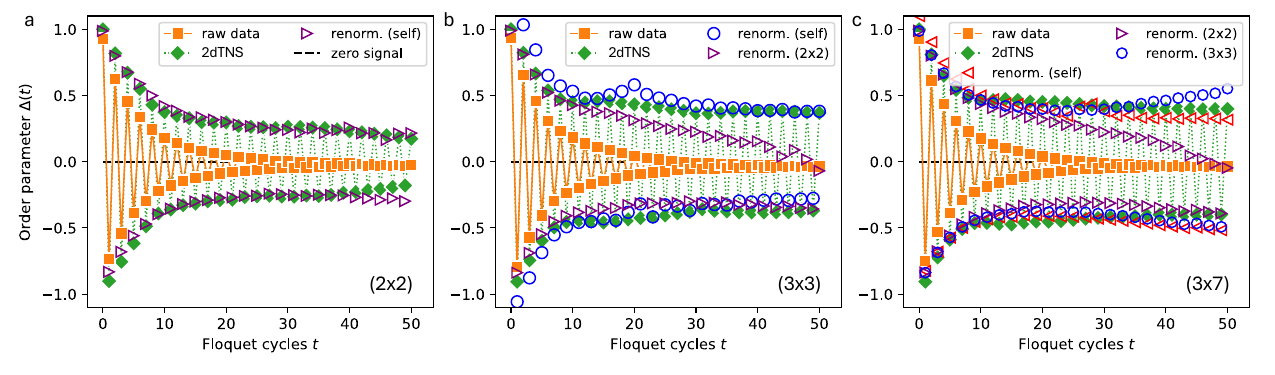}
    \caption{\textbf{Renormalization of the order parameter.} Comparison between experimental raw data, numerical simulations (2dTNS with bond dimension $\chi = 128$), and renormalized data. \textbf{a}, Time dependence of the $\Delta(t)$ parameter for $2\times2$ system, measured experimentally using Eq.~\eqref{eq:noisy_Delta} (orange, squares), computed numerically using the exact formula in Eq.~\eqref{eqs:delta_dfn} (green, diamonds), and renormalized using Eq.~\eqref{eq:polrz_recovered} (purple, right triangles). \textbf{b}, The same spin ordering parameter for a $3\times 3$ system. The normalized curve shows recovered data using Eq.~\eqref{eq:polrz_recovered} with parameters learned from a $2\times2$ system (purple, right triangles) and from the system itself (blue, circles). \textbf{c}, The same ordering parameters for a $3\times 7$ system. Renormalized curves represent recovered data using Eq.~\eqref{eq:polrz_recovered} with parameters learned from a $2\times2$ system (purple, right triangles), a $3\times3$ system (blue, circles), and the system itself (red, left triangles).}
  \label{fig:delta_mit_analysis}
\end{figure}

In the present series of experiments, the results obtained from the optimization of the expression in Eq.~\eqref{eqs:optim_J} for a classically simulable, smaller qubit subset (e.g., $2 \times 2$) are utilized to mitigate the results for a larger qubit subset (e.g., $3 \times 7$), where reliable classical simulation is not feasible. 
The plots shown in Figs.~\ref{fig:fig2}a and \ref{fig:polarized_state}b are generated using this approach. More data are shown in Fig.~\ref{fig:delta_mit_analysis}. 
The performance of the proposed  method is evaluated using examples of systems of sizes $2 \times 2$ (35 qubits) and $3 \times 3$ (68 qubits), see Fig.~\ref{fig:fez_2x2_3x3_qubits} and \ref{fig:fez_2x2_3x3_qubits_50cycle}, as well as $3 \times 7$ (144 qubits, see Fig.~\ref{fig:main_scheme}b): the vector of offsets $\vec{\delta}$ is learned by minimizing $J$ with $\Delta_{\mathrm{sim}}(\epsilon,\phi,t)$ simulated on $2 \times 2$ with parameters $\epsilon = 0.05$ and $\phi = 0.45\pi$.  
Then this offset is used to compute $\hat\Delta(0.05,0.45,t)$ for $3\times 3$, see Fig.~\ref{fig:delta_mit_analysis} (middle panel), and for $3\times7$, see Fig.~\ref{fig:delta_mit_analysis} (right panel) for 40 cycles.
However, this method has certain limitations. Specifically, if the learned values of $\delta_k(0, \phi_0, t)$ and $\delta_k(\epsilon, \phi_0, t)$ for the smaller subset differ significantly from their counterparts for the larger subset, this can lead to a scenario where the denominator of Eq.~\eqref{eq:polrz_recovered} is underestimated or even approaches zero. 
This discrepancy also explains the noticeable deviation in Fig.~\ref{fig:delta_mit_analysis} or signal increase observed at late times Figs.~\ref{fig:fig2}a and \ref{fig:polarized_state}b.

\subsection{Correlations}
\label{sec:supp_signalrecov_correlations}
A similar approach can be employed to mitigate the spin-spin ordering parameter. 
The noiseless spin-spin ordering parameter, defined over the set of spin pairs $S$, is given by  
    \begin{equation}
    \chi = \frac{1}{|S|}\sum_{(i,j)\in S}\langle Z_iZ_j\rangle^2\equiv \mathbb{E}_{S} \langle Z_iZ_j\rangle^2,
    \end{equation}  
where $|S|$ denotes the size of the set $S$, and $\mathbb{E}_{S}(\cdot) := \frac{1}{|S|} \sum_{(i,j)\in S}(\cdot)$ represents the expectation value over the set $S$.  In contrast, the noisy counterpart of this parameter can be expressed as  
    \begin{equation}
    \chi_{\rm noisy} := \frac{1}{|S|}\sum_{(i,j)\in S}\langle Z_iZ_j\rangle_{\rm noisy}^2 \equiv \mathbb{E}_{S} \langle Z_iZ_j\rangle_{\rm noisy}^2,
    \end{equation}  
where $\langle Z_iZ_j\rangle_{\rm noisy}$ denotes the expectation value of the correlator, obtained after noisy evolution and imperfect measurements.  Analogous to the treatment of individual spin polarizations, the impact of noise in the dynamics can be represented as  
    \begin{equation}
    \langle Z_iZ_j\rangle_{\rm noisy} = \varphi_{ij}(t)\langle Z_iZ_j\rangle + \eta_{ij}(t),
    \end{equation}  
where the function $\varphi_{ij}(t)$ characterizes the depolarizing component of the noise, and $\eta_{ij}(t)$ accounts for the noise bias.  

The spin glass order parameter is then expressed as,
    \be
    \begin{split}
    \chi_{\rm noisy}(t) &:= \mathbb E_S \langle Z_iZ_j\rangle_{\rm noisy}^2 = \mathbb E_S\Bigl(\varphi_{ij}(t)\langle Z_iZ_j\rangle + \eta_{ij}(t)\Bigl)^2\\
    & = \Bigl(\mathbb E_S  \varphi^2_{ij}(t) \langle Z_iZ_j\rangle^2 + 2 \mathbb E_S  \varphi_{ij}(t)\eta_{ij}(t)\langle Z_iZ_j\rangle + \mathbb E_S  \eta^2_{ij}(t)\Bigl).
    \end{split}
    \ee
First, assuming that the \textit{noiseless} expectation values are statistically independent of the prefactor $\varphi_{ij}(t)$, we can decouple the correlation function and rewrite the first term in the sum as,
    \be\label{eq:approx1-zz-corr}
    \begin{split}
    &\mathbb E_S \Bigl[ \varphi^2_{ij}(t) \langle Z_iZ_j\rangle^2 \Bigl]\approx \mathbb E_S \bigl[ \varphi^2_{ij}(t)\bigl] \mathbb E_{S} \bigl[\langle Z_{i'}Z_{j'}\rangle^2\bigl] + O\Bigl(\frac{1}{\sqrt{|S|}}\Bigl)\\
    & \qquad \qquad \qquad \qquad \qquad = \mathbb E_S\bigl[\varphi^2_{ij}(t)\bigl] \chi(t)+ O\Bigl(\frac{1}{\sqrt{|S|}}\Bigl),
    \end{split}
    \ee
where we used the definition of the noiseless spin-spin ordering parameter. 
Similarly, assuming independence between $\varphi_{ij}(t)$ and $\eta_{ij}(t)$, we can rewrite,
    \be\label{eq:approx2-zz-corr}
    \begin{split}
    &\mathbb E_S  \Bigl[\varphi_{ij}(t)\langle Z_iZ_j\rangle \eta_{ij}(t)\Bigl] \approx \mathbb E_S \bigl[ \varphi_{ij}(t)\langle Z_iZ_j\rangle\bigl]\mathbb E_S \bigl[\eta_{ij}(t)\bigl]+ O\Bigl(\frac{1}{\sqrt{|S|}}\Bigl) \\
    & \qquad \qquad \qquad \qquad \qquad = C_{\rm noisy}(t) \mathbb E_S \bigl[\eta_{ij}(t)\bigl] - \Bigl(\mathbb E_S \bigl[\eta_{ij}(t)\bigl] \Bigl)^2+ O\Bigl(\frac{1}{\sqrt{|S|}}\Bigl),
    \end{split}
    \ee
where we have introduced a new quantity that can be obtained from the experiment,
    \be
    C_{\rm noisy}(t) := \frac{1}{|S|}\sum_{i\neq j}\langle Z_iZ_j\rangle_{\rm noisy} \equiv \mathbb E_S\langle Z_iZ_j\rangle_{\rm noisy}.
    \ee
Combining the expressions from Eqs.~\eqref{eq:approx1-zz-corr} and \eqref{eq:approx2-zz-corr}, we arrive at the approximation for the noisy SG order parameter,
    \be
    \chi_{\rm noisy}(t) = (N - 1) \Bigl(\mathbb E_S  \varphi^2_{ij}(t) \chi_{\rm noisy}(t) + 2 \mathbb E_S  \eta_{ij}(t)C_{\rm noisy}(t) + \mathbb E_S  \eta^2_{ij}(t)-\bigl(\mathbb E_S \eta_{ij}(t) \bigl)^2\Bigl).
    \ee
Restoring the explicit dependence of all parameters on the physical variables $\epsilon$ and $\phi$, this expression can be written in a more compact form as,
    \be\label{eq:shift_chi}
    \chi_{\rm noisy}(\epsilon, \phi, t) = \varphi(\epsilon, \phi, t)  \chi(\epsilon, \phi, t)  -c_1(\epsilon, \phi)  C_{\rm noisy}(\epsilon, \phi, t)  - c_2(\epsilon, \phi),
    \ee
using the following notations,
    \be
    c_1 (\epsilon, \phi):= - 2 \mathbb E_S  \eta_{ij}(t)\Bigl|_{\epsilon, \phi}, \qquad c_2(\epsilon, \phi) :=\Bigl[2\bigl(\mathbb E_S \eta_{ij}(t) \bigl)^2 -\mathbb E_S  \eta^2_{ij}(t)\Bigl]\Bigl|_{\epsilon, \phi}.
    \ee
As a final step, similar to the previous section, we neglect the dependence of the collective parameter $\varphi(\epsilon, \phi, t)$ on the physical parameters $\epsilon$ and $\phi$, and write it as,
    \be
    \varphi(\epsilon, \phi, t)  \approx \varphi(t).
    \ee
Using this property and the expression in Eq.~\eqref{eq:shift_chi} for parameters $\epsilon = 0$ and $\phi = \phi_0$, we obtain,
    \be
    \chi(\epsilon,\phi,t) = \frac{\chi_{\rm noisy}(\epsilon, \phi, t) + 2 c_1(\epsilon,\phi) C_{\rm noisy}(\epsilon, \phi,t) + (N-1)c_2(\epsilon,\phi)}{\chi_{\rm noisy}(0, \phi_0, t) + 2 c_1(0,\phi_0) C_{\rm noisy}(0, \phi_0, t) + (N-1)c_2(0,\phi_0)},
    \ee
where we have taken into account that $\chi(0,\phi_0,t) = 1$.

\subsection{Hamming distance}
\label{sec:supp_signalrecov_hamming}
\begin{figure}[t!]
    \centering
\includegraphics[scale=0.8
]{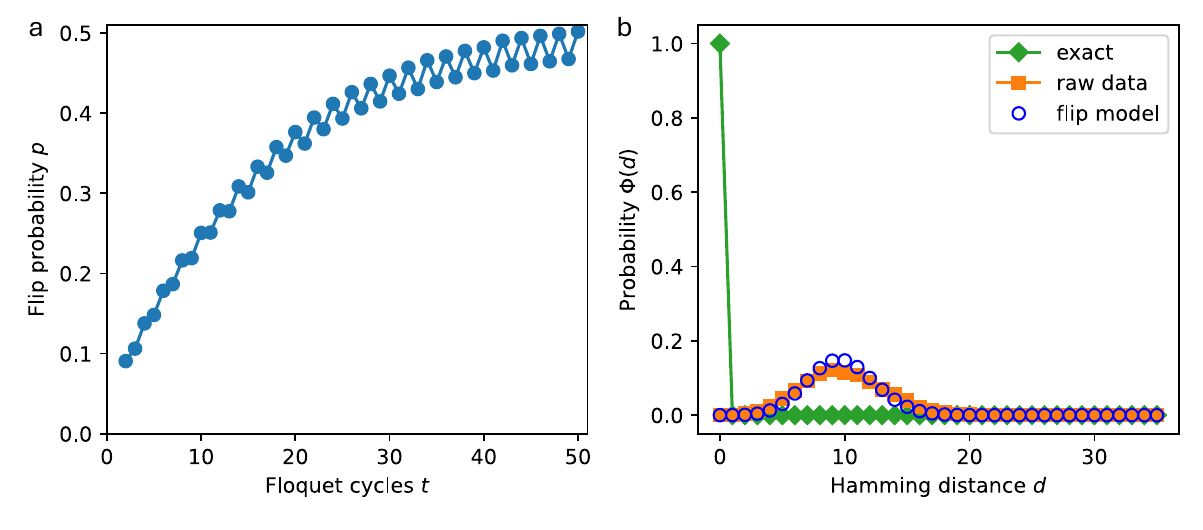}
    \caption{\textbf{Noise filtering for Hamming distance.} \textbf{a}, The learned flip parameter $p(t)$ as a function of the depth $t$ for the $2\times2$ system, obtained through the optimization procedure in Eq.~\eqref{eqs:finding_p_t}. \textbf{b}, Comparison of the experimental noisy distribution (orange, squares) with the distribution obtained by applying the kernel $T_p$ (blue, circles) to the true distribution (green, diamonds).}  \label{fig:ham_distr_filtering}
\end{figure}

We recall that the Hamming distance between the input state with polarizations ${\bf z} = \{s_i\}$ and the output string ${\bf z'} = \{s'_i\}$ is defined as,
    \be
    d({\bf z},{\bf z'}) := \frac 12\sum_{i=1}^N|s_i-s'_i| = \frac 12\sum_{i=1}^N(1-s_is'_i),
    \ee
which shows the number of spin flips required to transform one product state into the other. 
Correspondingly, the time-dependent noiseless distribution of Hamming distances can be defined,
    \be
    \Phi_d(t) = {\rm Prob}\bigl[d({\bf z}(t),{\bf z}(0)) = d\bigl].
    \label{eq:distr_ham_dist}
    \ee
The noisy distribution, obtained for the noisy output $z_{\rm noisy}(t) = \langle Z_i(t)\rangle_{\rm noisy}$, can be derived from the original noiseless distribution by applying a certain transformation represented by a time-dependent kernel $K(d,d'|t)$,
    \be
    \Phi^{\rm noisy}_d(t) := {\rm Prob}\bigl[d({\bf z}_{\rm noisy}(t),{\bf z}(0)) = d\bigl] = \sum_{d'=0}^N K(d,d'|t)\Phi_{d'}(t).
    \ee
While the exact form of this kernel is unknown and hard to derive, one could make an empirical approximation to it. 
One such approximation is to consider this kernel as a kernel $T_p(d,d')$ of a transformation that flips each spin independently with a certain time-dependent probability $p$, i.e.,
    \be
    K(d,d'|t) \approx T_p(d,d')\Bigl|_{p=p(t)}.
    \ee
A straightforward derivation provides an expression for this kernel in the form,
    \be
    T_p(d,d') = \sum^{\min(d,d')}_{\max(0,d+d'-N)} C^{d'}_x C^{n-d'}_{d-x}  p^{d+d'-2x} (1-p)^{N+2x-d-d'}.
    \ee
The parameter $p$ is unknown. 
It can be derived for a relevant experimental point if we neglect the dependence of $p(t)$ on the parameters $\epsilon$ and $\phi$. 
Then, this parameter can be learned from observing the Clifford point. This learning procedure is formulated as an optimization task,
    \be\label{eqs:finding_p_t}
    p(t) = {\rm argmin}[L_1(p,t)], \qquad L_1(p,t) = \sum_{d=0}^N\Bigl(\Phi^{\rm noisy}_d(t)\Bigl|_{\epsilon=0,\phi = \phi_0}- T_p\bigl(d,d_{\rm cliff}(t)\bigl)\Bigl)^2.
    \ee
The dependence of the learned value of $p(t)$ as a function of the depth $t$ is shown in Fig.~\ref{fig:ham_distr_filtering}a. 
The comparison between the experimental distribution and the distribution obtained using the kernel $T_p(d,d')$ is shown in Fig.~\ref{fig:ham_distr_filtering}b.

With the knowledge of the kernel $T_p(d,d')$, it is possible to ``reverse'' the effect of noise and learn the noiseless distribution. 
However, since the kernel has many eigenvalues that are close to zero, the outcome will depend significantly on small perturbations of the noisy distribution. 
To avoid this problem, we restrict the space of possible outputs by considering a fixed form of the noiseless distribution. 
In particular, we look only at distributions that take the form,
    \be
    \Phi^{\rm trial}_d = A\frac{\exp\left(-\frac{(d - d_0)^2}{2\sigma^2}\right)}{1 + \exp(kd + q)}
    \ee
where $d_0$, $\sigma$, $k$, and $q$ are free parameters, and $A$ is the normalization factor that depends on them.

We determine the most suitable trial distribution by minimizing the loss function,
    \be
    \Phi^{\rm trial}_d = {\rm argmin}[L_2],\qquad L_2 = \sum_{d=0}^N\Bigl(\Phi^{\rm noisy}_d(t)- \sum_{d'=0}^N T_p(d,d')\Phi^{\rm trial}_{d'}\Bigl)^2 + \lambda_1(\mu_{\rm trial}-\mu(t))^2+\lambda_2(\sigma_{\rm trial}-\sigma(t))^2,
    \ee
where $\lambda_1$ and $\lambda_2$ are positive parameters, $\mu_{\rm trial}$ and $\sigma_{\rm trial}$ are the mean value and variance of the distribution $\Phi^{\rm trial}_d$, while $\mu(t)$ and $\sigma(t)$ are the best estimates for the mean value and variance of the true noiseless distribution. 
The mean value is connected to the order parameter as,
    \be
    \mu(t) := \Bigl\langle\frac 12 \sum_i \Bigl(1-s_i(0)Z_i(t)\Bigl)\Bigl\rangle = \frac{N}{2}\Bigl(1-\Delta(t)\Bigl),
    \ee
where $\Delta(t)$ is the order parameter defined above in Eq.~\eqref{eqs:delta_dfn}. 
The variance, in turn, can be expressed as,
    \be
    \begin{split}
    \sigma(t) &:= \frac 14  \Bigl\langle\Bigl(\sum_i s_i(0)\bigl[Z_i(t)- \langle Z_i(t)\rangle\bigl]\Bigl)^2\Bigl\rangle = \frac 14 \sum_{ij} s_i(0)z_j(0)\Bigl(\langle Z_i(t)Z_j(t)\rangle-\langle Z_i(t)\rangle \langle Z_j(t)\rangle\Bigl).
    \end{split}
    \ee
This expression represents the quantum Fisher information.

\section{Edwards-Anderson spin glass order parameter}
\label{sec:supp_spinglass}
\begin{figure}[t!]
    \centering
\includegraphics[scale=0.45
]{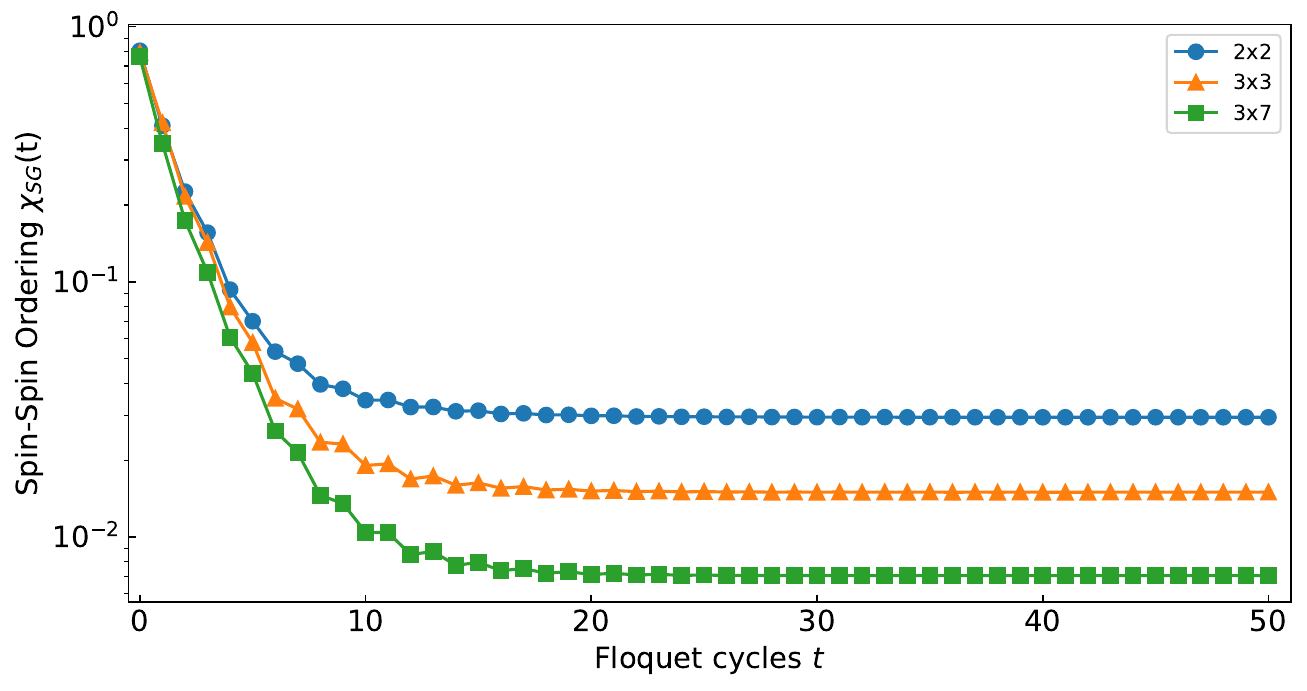}
    \caption{Comparison of the Edwards-Anderson spin glass order parameter derived from experimental raw data for the $2\cross 2$, $3\cross 3$, and $3\cross 7$ geometries at $\phi=0.45\pi$, $\epsilon=0.05$, demonstrating similar behavior to Fig.~\ref{fig:fig2}b.}
  \label{fig:chisg_allpairs}
\end{figure}

As described in the main text, if the set of spin pairs $M$ in Eq.~\ref{eq:spin_glass_order} is expanded to all qubit pairs, one recovers the Edwards-Anderson spin glass order parameter $\chi_{SG}$. 
The unmitigated $\chi_{SG}$ is shown in Fig.~\ref{fig:chisg_allpairs}, showing a clear similarity to that of the results of Fig.~\ref{fig:fig2}b.
\end{document}